\magnification1200


\vskip 2cm
\centerline
{\bf  Gravity,  Dual Gravity and   $A_1^{+++}$}
\vskip 1cm
\centerline{Keith Glennon and Peter West}
\centerline{Department of Mathematics}
\centerline{King's College, London WC2R 2LS, UK}
\vskip 2cm
\leftline{\sl Abstract}
We construct the non-linear realisation of the semi-direct product of the very extended algebra  $A_1^{+++}$ and its vector representation. This theory has an infinite number of fields that depend on a spacetime with an infinite number  of coordinates. Discarding all except the lowest level field and coordinates the dynamics is just Einstein's equation for the graviton field. We show that the gravity field is related to the dual graviton field by a duality relation and we also derive the equation of motion for the dual gravity field. 
\vskip2cm
\noindent

\vskip .5cm

\vfill
\eject

\medskip
{{\bf 1. Introduction}}
\medskip
Some time ago it was conjectured  that the non-linear realisation of the semi-direct product of $E_{11}$ and it's vector representation $(l_1)$, denoted $E_{11} \otimes_s l_1$, leads to the low energy effective action for the theory of strings and branes [1,2]. This theory contains an infinite number of fields associated with $E_{11}$ that live on a space-time that contains an infinite number of coordinates. The field equations follow from the symmetries of the non-linear realisation. If one takes the decomposition of $E_{11}$ into its GL(11) subalgebra then one finds a theory whose lowest level fields are those of eleven dimensional supergravity and the  level zero coordinates are those of the eleven dimensional spacetime we are familiar with. The essentially unique equation of motion that follow in this decomposition were found relatively recently
[3,4] and, in the restriction just mentioned,  they were precisely those  of eleven dimensional supergravity. To be more precise they were the equations of motion for the graviton $h_{a}{}^b$ and the three form field $A_{a_1a_2a_3}$.
\par
At the next two levels one finds a six form field $A_{a_1\ldots a_6}$ and a  field  $h_{a_1\ldots a_8, b}$. The former field was the well known dual of the three form while the latter fields was proposed to be the dual of the usual gravity field [1], the dual graviton. The field $h_{a_1a_2 ,b}$ had been previously investigated in five dimensions and proposed as a candidate for the dual graviton  [5],  while the  field $h_{a_1\ldots a_{D-3}, b}$ had been proposed in $D$ dimensions [6] as a candidate for the dual graviton. It was shown in reference [1] that this field did indeed describe the degrees of freedom of gravity in $D$ dimensions at the linearised level. as well as the references they contain. Previous work on the dual graviton in the context of $E_{11}$ can be found in references [7] and [8] and a review of E theory can be found in references [10],  [11] and [12]. 
\par
$E_{11}$ is a very extended algebra Kac-Moody algebra which can be found by adding three nodes to the Dynkin diagram of $E_{8}$ [13]. In terms of this construction one can write $E_{11}= E_8^{+++}$. Indeed this is a general procedure and one can add three nodes in this way to any semi-simple finite dimensional Lie Algebra, that is, the Lie algebras in the list of Cartan which was actually found by Killing. For each of these algebras  one can carry out a corresponding non-linear realisation. It was realised that for $K_{27}\equiv D_{24}^{+++}\otimes_s l_1$, where $l_1$ is the vector (first fundamental)  representation,  one finds the fields of the effective action of the twenty six dimensional bosonic string [1]. It is inevitable that the equations of motion of the lowest level fields are those of this effective action. It was also proposed that the very extended $A_{D-3}$ algebra, denoted by $A_{D-3}^{+++}$ describes gravity in $D$ dimensions [14]. These theories contain at lowest level the graviton field and at the next level the dual graviton  [15]. 
\par
In this paper we will consider the case of four dimensional gravity and so the non-linear realisation of algebra $A_{1}^{+++}\otimes_s l_1$ where this denotes the semi-direct product of $A_{1}^{+++}$ and it first fundamental (vector) representation $l_1$. The algebra
$A_{1}^{+++}\otimes_s l_1$ was worked out at low levels in reference [16] and the invariant tangent space metric and an invariant gauge fixing was found in reference [17]. In this paper we calculate the low level equations of motion for   the non-linear realisation of $A_1^{+++}\otimes_s l_1$  at low levels. If we restrict the theory to just contain the gravity
$h_a{}^b$ and dual gravity $\tilde h_a{}^b$ fields and the level zero coordinates $x^\mu, \mu=0,1,2,3$ then we find that the gravity field does obey Einstein's equation and a duality relation that relates the  gravity field to the  dual gravity field. We also derive the much sort after fully non-linear equation of motion for the dual gravity field. This equation involves the usual graviton field as well as the dual graviton and as a result it avoids the no go theoems of reference [9]. We also comment in the paper on the dual graviton equation derived in reference [8]. 
\par
The idea that gravity could be found as a result of a non-linear realisation dates back to an old paper of Aleksandr Borisov and Victor Ogievetsky [18] who proposed that gravity was the non-linear realisation of $GL(4)\otimes_s l_1$ where $l_1$ is the familiar vector representation. This non-linear realisation lead to equations of motion that were far from unique but they  proposed that one could take the simultaneous non-linear realisation of this algebra with the conformal algebra. This did lead uniquely to Einstein's equation as must have been the case as it was shown that the simultaneous action of GL(4) and the conformal group on the vector representation lead to general coordinate transformations [19]. In early days of $E_{11}$ when only some of the symmetries were being used to find the equations of motion it was proposed to also use the conformal group but this was found not to be helpful.
The uniqueness of the equations of motion  in the $E_{11} \otimes_s l_1$ non-linear realisation,  including that of gravity, was found to be a consequence of the higher level symmetries and in particular the local symmetries of the Cartan involution invariant subalgebra of $E_{11}$, denoted $I_c(E_{11})$, beyond those at level zero [3,4]. In this paper we will find that the equations of motion of gravity and dual gravity are essentially unique once we use the higher level symmetries in $A_{1}^{+++}\otimes_s l_1$.


\medskip
{\bf 2 The Kac-Moody algebra $A_1^{+++}$}
\medskip
We now establish the basic properties of the Kac-Moody algebra $A_1^{+++}$ and it's $l_1$ representation [16] at low levels. The Dynkin diagram for the Kac-Moody algebra $A_1^{+++}$ is 
$$
\matrix{
\bullet & - & \bullet & - & \bullet & = & \otimes \cr
1 & & 2 & & 3 & & 4 \cr
}
$$
which corresponds to the Cartan matrix
$$
A = \left(\matrix{
2 & -1 & 0 & 0 \cr
-1 & 2 & -1 & 0 \cr
0 & -1 & 2 & -2 \cr
0 & 0 & -2 & 2 \cr
}\right).
\eqno(2.1)$$
Like all Kac-Moody algebras that are not finite dimensional or affine there is no known listing of  the generators that $A_1^{+++}$. Deleting node four in the above  Dynkin diagram we find the residual algebra of $GL(4)$  and one can investigate the $A_1^{+++}$ algebra once it has been decomposed into this later algebra. The generators that one finds are classified by a level which is the number of up minus down $GL(4)$ indices on that generator all divided by two. The decomposition of $A_1^{+++}$ in terms of this subalgebra was  given at low levels [16].
\par
The positive level generators to low levels are given by
$$
K^a{}_b (16)  ; \quad R^{(ab)} (10); \quad  R^{a_1a_2,(b_1b_2)} (45)  ; \quad R^{a_1 a_2,b_1 b_2,(c_1 c_2)} (126)  , \quad  R^{a_1 a_2 a_3,b_1 b_2,c} (64)  ; $$
$$
R^{a_1 a_2,b_1 b_2, c_1 c_2,(d_1 d_2)}  , \quad R^{a_1 a_2 a_3,b_1 b_2,(c_1 c_2 c_3)} , \quad R^{a_1 a_2 a_3,b_1 b_2,c_1 c_2,d}_{(1)} , \quad R^{a_1 a_2 a_3,b_1 b_2,c_1 c_2,d}_{(2)} ,
$$
$$
R^{a_1 a_2 a_3,b_1 b_2 b_3,(c_1 c_2)} , \quad R^{a_1 a_2,(b_1 b_2)}, \quad R^{a_1 a_2 a_3,d}, \ldots
\eqno(2.2)$$
where the generators at levels zero, one, two, ... are separated by a semi-colon and the numbers in brackets for the first few generators are the dimensions of the representations. All the upper indices are assumed to be anti-symmetric except for the indices which appear with  $( \ )$ brackets and these are symmetric. In what follows  we will drop these brackets,  for example $R^{[a_1 a_2],(b_1 b_2)}$ will just be written as $R^{a_1 a_2,b_1b_2}$. The  subscript  indicate that a   generator has multiplicity greater than one and the   $(1)$ and $(2)$ distinguishing the different generators. These generators possess the $GL(4)$ irreducibility properties
$$
R^{[a_1a_2,b_1]b_2} = 0, \ \ R^{[a_1 a_2, b_1] b_2,c_1 c_2} = 0, \ \ \ R^{[a_1 a_2,|b_1b_2|,c_1]c_2} = 0,$$
$$
R^{[a_1 a_2 a_3,b_1]b_2,c} = 0, \ \ \ R^{a_1 a_2 a_3,[b_1b_2,c]} = 0, \ldots 
\eqno(2.3)$$
The negative level generators $R_{ab}, \ R_{ab,cd}, \ldots $ possess analogous symmetry and irreducibility properties to their positive level counterparts.
\par
The generators belong to representations of $GL\left(4\right)$ and so the commutators of  $K^a{}_b$ with the positive generators are
$$
\left[K^a{}_b,\,K^c{}_d \right] = \delta^{c}{}_{b}\,K^a{}_d - \delta^{a}_{d}\,K^c{}_b,
$$
$$ \left[K^a{}_b,\,R^{c_1c_2} \right] = 2\,\delta^{(c_1}_{\,b}\,R^{|a|c_2)},\quad \left[K^a{}_b,\,R_{c_1c_2} \right] = -\,2\,\delta^{\,a}_{(c_1}\,R_{|b|c_2)},
$$
$$
\left[K^a{}_b,\,R^{cd,ef} \right] = \delta^{c}_{b}\,R^{ad,ef} + \delta^{d}_{b}\,R^{ca,ef} + \delta^{e}_{b}\,R^{cd,af} + \delta^{f}_{b}\,R^{cd,ea},
$$
$$
\left[K^a{}_b,\,R_{cd,ef} \right] = -\,\delta_{c}^{a}\,R_{bd,ef} - \delta_{d}^{a}\,R_{cb,ef} - \delta_{e}^{a}\,R_{cd,bf} - \delta_{f}^{a}\,R_{cd,eb}. \eqno(2.4)
$$
$$
[K^a{}_b,R^{c_1 c_2,d_1 d_2,e_1 e_2}] = \delta^{c_1}{}_b R^{a c_2,d_1 d_2,e_1 e_2} + ... + \delta^{e_2}{}_b R^{c_1 c_2,d_1 d_2,e_1 a} ,$$
$$
[K^a{}_b,R_{c_1 c_2,d_1 d_2,e_1 e_2}] = - \delta^{a}{}_{c_1} R_{b c_2,d_1 d_2,e_1 e_2} - ... - \delta^{a}{}_{e_2} R_{c_1 c_2,d_1 d_2,e_1 b} , $$
$$
[K^a{}_b,R^{c_1 c_2 c_3,d_1 d_2,e}] = \delta^{c_1}{}_b R^{a c_2 c_3,d_1 d_2,e} + \delta^{c_2}{}_b R^{c_1 a c_3,d_1 d_2,e} + ... + \delta^{e}{}_b R^{c_1 c_2 c_3,d_1 d_2,a},$$
$$
[K^a{}_b,R_{c_1 c_2 c_3,d_1 d_2,e}] = - \delta^{a}{}_{c_1} R_{b c_2 c_3,d_1 d_2,e} - ... - \delta^{a}{}_e R_{c_1 c_2 c_3,d_1 d_2,b},$$
\par
The commutators of the  level 2 ($-2$)  must give on the right-hand side the unique level 2 ($-2$) generators and so these commutators must be of the form
$$
\left[ R^{ab},\,R^{cd} \right] = R^{ac,bd} + R^{bd,ac}, \quad \left[ R_{ab},\,R_{cd} \right] = R_{ac,bd} + R_{bd,ac}. \eqno(2.5)
$$
where the normalisation of the level 2 ($-2$) generators are fixed by these relations. The commutators between the positive and negative level generators are given by
$$
\left[R^{ab},\,R_{cd}\right] = 2\,\delta^{(a}_{(c}\,K^{b)}{}_{d)} - \delta^{(ab)}_{cd}\sum_eK^{e}{}_{e}, $$
$$
\left[R^{ab,cd},\,R_{ef}\right] = \delta_{ef}^{(bd)}\,R^{ac} + \delta_{ef}^{(bc)}\,R^{ad} - \delta_{ef}^{(ac)}\,R^{bd} - \delta_{ef}^{(ad)}\,R^{bc},
$$
$$
\left[R_{ab,cd},\,R^{ef}\right] = \delta^{ef}_{bd}\,R_{ac} + \delta^{(ef)}_{bc}\,R_{ad} - \delta^{(ef)}_{ac}\,R_{bd} - \delta^{ef}_{ad}\,R_{bc}. \eqno(2.6)
$$
where $\delta^{(ab)}_{cd}= \delta ^{(a}_c\delta ^{b)}_d$.
\par
The Cartan involution acts on the generators of $A_1^{+++}$ as follows 
$$
I_c\left(K^a{}_b\right) = -\,K^b{}_a, \quad I_c\left(R_{ab}\right) = -\,R^{ab}, \quad
I_c\left(R^{ab,cd}\right) = R_{ab,cd}, \ldots .
\eqno(2.7)$$
The Cartan-involution invariant generators are given by
$$
J_{ab} = \eta_{ac} K^c{}_b - \eta_{bc} K^c{}_a, \ \
$$
$$
S_{ab} = R^{cd} \eta_{ca} \eta_{db} - R_{ab}, \ \ S_{a_1 a_2,b_1 b_2} = R^{c_1 c_2,d_1 d_2} \eta_{c_1 a_1} \eta_{c_2 a_2} \eta_{d_1 b_1} \eta_{d_2 b_2} - R_{a_1 a_2,b_1b_2} ,\ldots
\eqno(2.8)$$
\par
They  generate the Cartan involution-invariant subalgebra denoted by $I_c(A_1^{+++})$ whose low level commutators are
$$
[J_{a_1 a_2},J_{b_1 b_2}] = \eta_{a_2 b_1} J_{a_1 b_2} - \eta_{a_2 b_2} J_{a_1 b_1} - \eta_{a_1 b_1} J_{a_2 b_2} + \eta_{a_1 b_2} J_{a_2 b_1}$$
$$
[J_{a_1 a_2},S_{b_1 b_2}] = \eta_{a_2 b_1} S_{a_1 b_2} + \eta_{a_2 b_2} S_{a_1 b_1} - \eta_{a_1 b_1} S_{a_2 b_2} - \eta_{a_1 b_2} S_{a_2 b_1}$$
$$
[S_{a_1 a_2},S_{b_1 b_2}] = 2 S_{(a_1 |(b_1,b_2)|a_2)} - 2 \eta_{( b_1|(a_1 }  J_{a_2)| b_2)} , \ \ \ldots \eqno(2.9)$$
\par
The first fundamental representation, also called the  vector representation,  is denoted by $l_1$. This  representation has,  at low levels, the generators
$$
P_a \ ; \ \ Z^a \ ; \ \ Z^{(a_1 a_2 a_3)}, \ \ Z^{a_1 a_2,b}, \ \  Z^{a_1 a_2,(b_1 b_2 b_3)}, \ \ Z^{a_1a_2,b_1 b_2,c}_{(1)}, \ \ Z^{a_1a_2,b_1 b_2,c}_{(2)}, \ \ $$
$$
Z^{a_1 a_2 a_3,(b_1 b_2)}, \ \ Z^{a_1a_2a_3,b_1b_2}, \ \ Z^{a_1a_2,b_1b_2,(c_1c_2c_3)}_{(1)}, \ \ Z^{a_1a_2,b_1b_2,(c_1c_2c_3)}_{(2)}, \ \ \ldots   \eqno(2.10)
$$
where, as before,  the upper indices with no brackets are anti-symmetric, while those with $( \ )$ brackets are symmetric. The subscripts denote the different generators when the multiplicity is greater than one.  These generators satisfy the irreducibility conditions
$$
Z^{[a_1 a_2,b]} = 0, \ \ ... \eqno(2.11)$$
\par
The semi-direct product of the  $A_1^{+++}$ with the generators in $l_1$ representation is denoted by  $A_1^{+++}\otimes_s l_1$. The commutators of the   $A_1^{+++}$ generators with those of the vector representation have the form
$$
[K^a{}_b,\,P_c] = -\,\delta^a_c\,P_b + {1\over 2}\,\delta^a_b\,P_c, \quad [K^a{}_b,\,Z^c] = \delta^c_b\,Z^a + {1\over 2}\,\delta^a_b\,Z^c,
$$
$$
[K^a{}_b,\,Z^{cde}] = \delta^c_b\,Z^{ade} + \delta^d_b\,Z^{cae} + \delta^e_b\,Z^{cda}+{1\over 2} \delta_b^a Z^{cde} ,
$$
$$
[K^a{}_b,\,Z^{cd,e}] = \delta^c_b\,Z^{ad,e} + \delta^d_b\,Z^{ca,e} + \delta^e_b\,Z^{cd,a}+{1\over 2} \delta_b^a Z^{cd, e}. \eqno(2.12)
$$
$$
[R^{ab},\,P_c] = \delta^{(a}_{\,c}\,Z^{b)}, \quad [R^{ab},\,Z^{c}] = Z^{abc} + Z^{c(a,b)}.
$$
$$
[R^{ab,cd},\,P_e] = -\,\delta^{[a}_{\,e}\,Z^{b]cd} + {1\over 4}\,\left( \delta^a_e\,Z^{b(c,d)} - \delta^b_e\,Z^{a(c,d)} \right) - {3\over 8}\,\left( \delta^c_e\,Z^{ab,d} + \delta^d_e\,Z^{ab,c} \right)
$$
The commutators with the  negative level $A_1^{+++}$ generators are given by
$$
[R_{ab},\,P_c] = 0, \quad [R_{ab},\,Z^c] = 2\,\delta^{\,c}_{(a}\,P_{b)},
$$
$$
[R_{ab},\,Z^{cde}] = {2\over 3}\,\left( \delta^{cd}_{(ab)}\,Z^e + \delta^{de}_{(ab)}\,Z^c + \delta^{ec}_{(ab)}\,Z^d \right),
$$
$$
[R_{ab},\,Z^{cd,e}] = {4\over 3}\,\left( \delta^{de}_{(ab)}\,Z^c - \delta^{ce}_{(ab)}\,Z^d \right). \eqno(2.13)
$$
\par
We also have an algebra formed from the $I_c(A_1^{+++})$ generators and  the $l_1$ generators
$$
[J_{a_1 a_2},P_b] = 2 P_{[a_1} \eta_{a_2]b}, \ \ \  [S^{ab},P_c] = \delta^{(a}{}_c Z^{b)}, $$
$$
[J^{a_1 a_2},Z^b] = - 2 \eta^{b[a_1} Z^{a_2]},$$
$$
[S^{a_1 a_2},Z^b] = ( Z^{a_1 a_2 b} + Z^{b(a_1,a_2)} )  - 2 \eta^{b (a_1} P^{a_2)}. \eqno(2.14) $$


\medskip
{\bf{3. Non-linear realisations of  $A_1^{+++} \otimes_s l_1$}}
\medskip
The construction of the non-linear realisation of $E_{11}\otimes_s l_1$ was  discussed in detail in the previous papers on $E_{11}$. The reader may like to look at reference [10] and the review of reference [11].  The general features of this construction apply to the non-linear realisation of $A_1^{+++} \otimes_s l_1$ which  we now briefly summarise. It starts with the group element group element $g\in A_1^{+++} \otimes_s l_1$ that can be written as
$$
g=g_lg_A
\eqno(3.1)$$
In this equation $g_A$ is a group element of $A_1^{+++}$ which can  be written in the form
$g_A= \Pi_{\underline \alpha} e^{A_{\underline \alpha} R^{\underline \alpha}}$ where the $R^{\underline \alpha}$ are  the  generators of $A_1^{+++}$ given in equations (2.2) as well as their negative level counter parts. The group element $g_l$ is formed from the generators of the vector ($l_1$) representation and so has the form $\Pi_A e^{z^A L_A} $ where $z^A$ are the coordinates of the generalised space-time. The fields $A_{\underline\alpha}$ depend on the coordinates $z^A$. 
\par
The above group elements can, up to level three,  be written in the form
$$
g_A = \ldots\,e^{A_{a_1a_1a_3,b_1b_2,c}R^{a_1a_2a_3,b_1b_2,c}}
$$
$$
e^{A_{a_1a_2,b_1b_2,c_1c_2}\,R^{a_1a_2,b_1b_2,c_1c_2}}\,e^{A_{a_1a_2,b_1b_2}\,R^{a_1a_2,b_1b_2}}\,e^{A_{a_1a_2}\,R^{a_1a_2}} \, e^{h_{a}{}^{b}\,K^{a}{}_{b}} \ldots \, .
\eqno(3.2)$$
where $\dots $ at the beginning of the equation corresponds to the presence of the higher positive level generators and the
$\dots $ at the end of  the equation corresponds to the presence of the negative  level generators
While the group element $g_l$ can be taken to be of  the form
$$
g_l = e^{x^a\,P_a}\,e^{y_a\,Z^a}\,e^{x_{abc}\,Z^{abc}}\,e^{x_{ab,c}\,Z^{ab, c}}\ldots
\eqno(3.3)$$
\par
In the above group elements we have introduced the fields 
$$
h_{a}{}^{b}; \  A_{(a_1 a_2)}; \ A_{a_1a_2,(b_1b_2)} ; \ A_{a_1a_2,b_1b_2,(c_1c_2)} \  A_{a_1a_2a_3,b_1b_2,c}, 
$$
$$ 
 A_{a_1a_2, b_1b_2, c_1c_2, (d_1d_2)} , \ A_{a_1a_2a_3, b_1b_2, c_1c_2, d}, \ldots
\eqno(3.4)$$
where as a block of indices is antisymmetric in its indices, except if if it is contained between $()$ in which case it is symmetrised. We will in what follows drop these latter brackets but the reader should recall that the indices are symmetrised. The fields obey the GL(4) irreducibility conditions, for example 
$$A_{[a_1a_2, b_1]b_2}=0, \ A_{[a_1a_2, b_1]b_2, c_1c_2}=0,
$$
$$ \ A_{[a_1a_2|, b_1b_2, |c_1]c_2}=0, \ 
A_{[a_1a_2a_3, b_1]b_2, c}=0, \ A_{a_1a_2, [b_1b_2, c_1]c_2}=0, \ 
\eqno(3.5)$$
These fields have 45, 126 and 64 components respectively. In arriving at this count we took account of the fact that $A_{[a_1a_2, b_1b_2]}=0$ as well as similar conditions for the other two fields. 
\par
We have also introduced the  generalized coordinates of the  space-time
$$
x^a; \ y_a; \ x_{abc}, \ x_{ab,c}; x_{a_1a_2,b_1b_2b_3}, \ x_{a_1a_2,b_1b_2,c}, \ x_{a_1a_2a_3,b_1b_2}, \   x_{a_1a_2a_3,(b_1b_2)}, \ \ldots
\eqno(3.6)$$
which possess the same symmetries as their corresponding generators in the vector representation, for example $x_{abc}= x_{(a_1 a_2 a_3)}$.
The fields and coordinates obey the same irreducibility  as their corresponding generators.
\par
The field $h_{a}{}^{b}$ is the usual graviton,  the field $ A_{ab}$ is the dual graviton and the field $A_{ab,cd}$ is the dual dual-graviton etc. The coordinates $x^a$ are the usual coordinates of space-time while the coordinates $y_a$ are the coordinates associated with the dual graviton. expand.
\par
The non-linear realisation is, by definition, invariant under the transformations
$$
g\to g_0 g, \ \ \ g_0\in A_1^{+++} \otimes_s l_1,\ \ {\rm as \  well \  as} \ \ \ g \to gh, \ \ \ h\in
I_c(A_1^{+++})
\eqno(3.7)$$
The group element $g_0 \in A_1^{+++}$ is a rigid transformation, that is, it is  a constant. The group element $h$ belongs to the Cartan involution invariant subalgebra $I_c(A_1^{+++})$ of $A_1^{+++}$ and  it is a local transformation meaning that  it depends on the coordinates of the  space-time.
\par
As the generators in $g_l$ form a representation of $A_1^{+++}$ the above transformations for $g_0\in A_1^{+++}$ can be written as
$$
g_l\to g_0 g_lg_0^{-1}, \quad g_A \to g_0 g_A\quad \ {\rm and}  \ \quad g_A \to g_A h
\eqno(3.8)$$
Using these transformations we can set to zero all parts of the group element $g_A$ which depend on the negative level generators.
\par
The dynamics of the non-linear realisation is just a set of equations of motion, that are invariant under the transformations of equation (3.7). We will  construct the dynamics of the $A_1^{+++} \otimes_s l_1$ non-linear realisation from the Cartan forms which are given by
$$
{\cal V} \equiv g^{-1} d g= {\cal V}_A+ {\cal V}_l,
\eqno(3.9)$$
where
$$
{\cal V}_A=g_A^{-1}dg_A \equiv dz^\Pi G_{\Pi, \underline \alpha} R^{\underline \alpha}, \  \ {\rm and}  \
{\cal V}_l= g_A^{-1}(g_l^{-1}dg_l) g_A = g_A^{-1} dz \cdot l g_A \equiv
dz^\Pi E_\Pi{}^A l_A 
\eqno(3.10)$$
Clearly ${\cal V}_A$ belongs to the $A_1^{+++}$ algebra and it is  the Cartan form of $A_1^{+++}$ while ${\cal V}_l$ is in the space of generators of the $l_1$ representation. The object  ${E}_{\Pi}{}^A = (\Pi_{\alpha} e^{A_{\underline \alpha}D^{\underline \alpha}})_{\Pi}{}^A$ is the vielbein on the  spacetime introduced in the non-linear realisation.
\par
Both ${\cal V}_A$ and ${\cal V}_l$, when viewed as forms,  are invariant under rigid transformations,  but  under the local $I_c(A_1^{+++})$ transformations of equation (1.3) they change as
$$
{\cal V}_A \to h^{-1}{\cal V}_A h + h^{-1} d h\quad \ {\rm and} \quad
{\cal V}_l\to h^{-1}{\cal V}_l h
\eqno(3.11)$$
\par
The Cartan form of  $I_c(A_1^{+++})$ can be written as
$$
{\cal V}_A = G_a{}^b K^a{}_b  + \overline{G}_{a_1 a_2} R^{a_1 a_2} + G_{a_1 a_2,b_1 b_2} R^{a_1 a_2 , b_1 b_2} + \ldots
\eqno(3.12)$$
Substituting the group element of equation (3.2) we find that the Cartan forms are given by
$$
G_a{}^b = (e^{-1} d e)_a{}^b $$
$$
\overline{G}_{a_1 a_2} = e_{a_1}{}^{\mu_1} e_{a_1}{}^{\mu_2} d A_{\mu_1 \mu_2}, $$
$$
G_{a_1 a_2,b_1 b_2} = e_{a_1}{}^{\mu_1} e_{a_2}{}^{\mu_2} e_{b_1}{}^{\nu_1}{} e_{b_2}{}^{\nu_2}(dA_{\mu_1 \mu_2,\nu_1 \nu_2} - A_{[\mu_1|(\nu_1} d A_{\nu_2)|\mu_2]}  )
\eqno(3.13)$$
One can easily  verify that $G_{a_1 a_2,b_1 b_2}$ really does satisfy the irreducibility condition $G_{[ a_1 a_2,b_1]  b_2}=0$.  The presence of the
$(\det e )^{{1\over 2}}$ factors arises from the unexpected terms with coefficient one half in equation (2.12).
\par
The generalised vielbein and it's inverse up  to level one [16] are given by
$$
E_\Pi{}^A = \left( \det{e} \right)^{-\,{1\over 2}}   \left(
\matrix{
e_\mu{}^a & - e_\mu{}^b A_{ba}  \cr
0 & \left( e^{-1} \right)_a{}^\mu \cr}
\right), \ \ \ \left(E^{-1} \right)_A{}^\Pi = \left( \det{e} \right)^{1\over 2}   \left(
\matrix{
\left( e^{-1} \right)_a{}^{\mu} & A_{ab} e_{\mu}{}^b  \cr
0 & e_{\mu}{}^a \cr}
\right), \eqno(3.14) $$
\par
The Cartan form transforms under the local $I_c(A_1^{+++})$ transformation as expressed in equation (3.11). The Cartan involution invariant subalgebra  at level zero is the Cartan involution invariant subalgebra of $GL(4)$ which  is $SO(1,3)$ and the Cartan forms transform under this  symmetry  as their indices suggest.  At the next level they transform under the group element $h = I - \Lambda_{a_1 a_2} S^{a_1 a_2} \in I_c(A_1^{+++})$ as
$$
\delta {\cal V}_A = [\Lambda_{a_1 a_2}S^{a_1 a_2},{\cal V}_A] - S^{a_1 a_2} d \Lambda_{a_1 a_2}
\eqno(3.15)
$$
These   variations are given explicitly by
$$
\delta G_a{}^b = 2 \Lambda^{cb} \overline{G}_{ca} - \delta_a{}^b \Lambda^{c_1 c_2} \overline{G}_{c_1 c_2} , \ \ \ \delta \overline{G}_{a_1 a_2} = - 2 \Lambda_{(a_1}{}^b G_{a_2) b} - 4 G_{(a_1 |b_1|, a_2) b_2} \Lambda^{b_1 b_2} - d \Lambda_{a_1 a_2} $$
$$
\delta G_{a_1 a_2,b_1 b_2} = 2 \Lambda_{[a_1 |(b_1|} G_{|a_2] | b_2)} ,
\eqno(3.16)
$$
\par
As in the $E_{11}$ case [4],  we must require that  the local transformations of equation (3.15) preserve the gauge choice. Demanding that the transformed Cartan form has no negative level parts we find that the $\Lambda_{a_1 a_2}$ parameter is restricted by
$$
d \Lambda_{a_1 a_2} - 2 \Lambda_{(a_1}{}^b G_{|b|a_2)} = 0.
\eqno(3.17)$$
This equation implies that the parameter $\Lambda^{\mu\nu}$, that is, the one with upper world indices,  is  a constant.
Using equation (3.17) in equation (3.16) we find that we can re-express $\delta \overline{G}_{a_1 a_2}$ as
$$
\delta \overline{G}_{a_1 a_2} = - 4 \Lambda_{(a_1}{}^b G_{(a_2) b)} - 4 G_{(a_1 |b_1|, a_2) b_2} \Lambda^{b_1 b_2}  .
\eqno(3.18)
$$
\par
While the Cartan forms when written as forms are invariant under the above transformations once we consider them as components, that is, we remove the forms $dz^\Pi$ they are no longer invariant under the rigid transformation $g_0\in A_1^{+++} \otimes_s l_1$. To get an object that is invariant under these rigid transformations we consider the objects
$$
G_{A,\bullet}= (E^{-1})_A{}^\Pi G_{\Pi , \bullet}
\eqno(3.19)$$
where $\bullet $ is any $A_1^{+++}$ index. However, these $A$ indices transform under the local $h\in I_c(A_1^{+++})$ transformations given by equation (3.11) and as a result on their first ($l_1$)  index the Cartan forms of equation (3.19) transform as
$$
\delta G_{a,\bullet} = - \Lambda_{ab} \hat{G}^b, \ \ \  \delta \hat{G}^{a,}{}_{\bullet} = 2 \Lambda^{ab} G_{b,\bullet}, 
\eqno(3.20)$$
where the hat indicates a derivative with respect to the level one coordinate $y_a$. Thus the  Cartan forms transform under the simultaneous effect of equations (3.16), (3.18) and (3.20).
\par
The non-linear realisation results in an invariant set of equations which are constructed from the  fields of the theory of equation (3.4) which depend on the generalised space-time coordinates of equation (3.6).


\medskip
{\bf{4. Derivation of the Duality Equations}}
\medskip
We will now construct equations that are first order in derivatives using the Cartan forms of equation (3.13) which  are invariant under the rigid
$g_0\in A_1^{+++} \otimes_s l_1$. As a result  we do not need to take further account of these transformations.   The Cartan forms do, however,  transform under the local $h\in I_c(A_1^{+++})$ transformations and so it invariance under these transformations that we will require.  At level zero the $I_c(A_1^{+++})$ transformations are just local Lorentz transformations $SO(1,3)$. While  the  transformations at the level one are given in equations (3.16) , (3.18) and (3.20). As for the case of $E_{11}$ we demand that  these first order equations will only be invariant under the  above  transformations of the non-linear realisation but modulo certain gauge transformations. This some what subtle point is explained in detail in references [20,4,12 ].
\par
The level one transformations of equations  (3.16) and  (3.18) transform Cartan forms of a given level into Cartan forms that have a level increased or decreased by one. Hence the  variation of  the Cartan form associated with our usual formulation of gravity, that is, the one constructed from  the gravity  field $h_a{}^b$, will   led to the Cartan form associated with the field  $\tilde A_{a_1a_2}$ associated with dual graviton. Thus we expect a duality relation that relates the gravity field to the dual gravity field.  We will  start by  considering the  well known spin connection which in terms of the gravity Cartan form is given by
$$
(\det e)^{1/2} \omega_{a,b_1 b_2}  =(- G_{b_1 , (b_2 a)} + G_{b_2, (b_1 a)} + G_{a,[b_1 b_2]}) 
\eqno(4.2)$$
While one could proceed by  writing  down the most general  equation constructed from the Cartan forms and test its invariance it is easier, and equivalent, to start from the spin connection and see what terms one must add by demanding $I_c(A_1^{+++})$ invariance. 
 \par
Simply using Lorentz symmetry we find that the equation should be of the generic form
$$
E_{a,b_1 b_2}  \equiv (\det e)^{1/2} \omega_{a,b_1 b_2} + {\tilde{e}_1 \over 2} \varepsilon_{b_1 b_2}{}^{c_1 c_2} \overline{G}_{c_1,c_2 a} \dot = 0
\eqno(4.3)
$$
where $\tilde{e}_1$ is a constant. The factor of $(\det e )^{{1\over 2}}$ correspond to the same factors in equation (3.13), that is, such factors  appear in the Cartan forms.
\par
We observe that the spin connection has to transform under local $I_c(A_1^{+++})$ transformations into not just the dual gravity Cartan form but the one which has its first two indices anti-symmetrised, namely $\overline{G}_{[c_1,c_2 ]a}$ as it is this object that occurs in equation (4.3). While the spin connection does not  do this we can add to it terms that involve derivatives with respect to higher level coordinates, the so called $l_1$ terms, such that it does. The required object is
$$
(\det e)^{1/2} \Omega_{a,b_1 b_2} = (\det e)^{1/2} \omega_{a,b_1 b_2} - {1 \over 2} \eta_{b_2 a}  \hat{\overline{G}}{}^{e,}{}_{b_1 e} + {1 \over 2} \eta_{b_1 a} \hat{\overline{G}}{}^{e,}{}_{b_2 e} \eqno(4.4)
$$
Its variation is given by 
$$
\delta [ (\det e)^{1/2} \Omega_{a,b_1 b_2} ] = 2 \Lambda^e{}_{a} \overline{G}_{[b_2, b_1] e}  + 2 \Lambda^e{}_{b_2} \overline{G}_{[a, b_1] e}   + 2 \Lambda^e{}_{b_1} \overline{G}_{[b_2,a] e} + 2 \eta_{b_2 a} \Lambda^{e_1 e_2} \overline{G}_{[b_1,e_1] e_2}  $$
$$
- 2 \eta_{b_1 a} \Lambda^{e_1 e_2} \overline{G}_{[b_2,e_1] e_2} .
\eqno(4.5) $$
In arriving at this result we have used equation  (3.20).
\par
As for the case of $E_{11}$, we will only  compute the equations of motion and duality relations to lowest level in the derivatives of the coordinates, meaning that they contain only derivatives with respect to the usual coordinates $x^a$ of spacetime. As a result we only keep terms in the local $I_c(A_1^{+++})$ variations that have no derivatives with respect to the higher level coordinates. However, terms in the equation that is being varied that are linear in derivatives with respect to the level one coordinates $y_a$ will, according to equation (3.20), vary into terms that have  ordinary derivatives. Such  terms will  contain  as one of its factors the Cartan forms $\hat G_{a,\bullet}$.  As a result we will require such terms in the equations we are varying. We will refer to such terms as  $l_1$ terms. 
\par
To summarise we will find the equation that is the result of the variation  only  up to  derivatives with respect to  the level zero coordinates but to do this we will  be required to find  the equations that are being varied up to derivatives  with respect to the level one coordinates. Indeed by varying equations one can find the terms that they contain that have  derivatives  with respect to the level one coordinates. We will refer to this as the $l_1$ extension of the equation. 
\par
Taking all this into account we vary the object $E_{a,b_1 b_2} $ of equation (4.3) but  as a help along the way we may use the object of equation (4.4) instead of  the usual spin connection. Adding further $l_1$ terms one finds that the  $l_1$ extended object duality relation between the gravity and dual gravity fields is given by 
$$
{\cal E}_{a,b_1 b_2} \equiv  (\det e)^{1/2} \Omega_{a,b_1 b_2}  + {1 \over 2} \varepsilon_{b_1 b_2}{}^{c_1 c_2} \overline{G}_{c_1,c_2 a} + {1 \over 2} \varepsilon_{b_1 b_2}{}^{c_1 c_2} (  \hat{G}_{c_2,[c_1 a]} + {1 \over 2}   \hat{G}_{a,[c_1 c_2]} ) $$
$$
- {1 \over 2} \varepsilon_{b_1 b_2}{}^{c_1 c_2} ( \hat{G}^{e ,}{}_{c_1 a, c_2 e} + {1 \over 2}  \hat{G}^{e,}{}_{c_1 c_2, a e})
- {1 \over 4}  \eta_{a b_1} \overline{G}^{e,}{}_{b_2 e} + {1 \over 4} \eta_{a b_2} \Lambda^{e_1 e_2}  \overline{G}^{e,}{}_{b_1 e}\dot =0
\eqno(4.6)$$
and it  varies under a  local $I_c(A_1^{+++})$ transformations as follows
$$
\delta {\cal E}_{a,b_1 b_2} = {1 \over 2} \varepsilon_{b_1 b_2}{}^{c_1 c_2} \Lambda_{a}{}^e E_{e,c_1 c_2} + \varepsilon_{b_1 b_2}{}^{c_1 c_2} \Lambda_{c_2}{}^e  E_{e,c_1 a}
- {1 \over 2} \eta_{a b_1} \Lambda^{e_1 e_2} \overline{E}_{b_2,e_1 e_2}
$$
$$ +  {1 \over 2} \eta_{a b_2} \Lambda^{e_1 e_2} \overline{E}_{b_1,e_1 e_2}  + \Lambda_{b_1}{}^e  \overline{E}_{b_2,ae} - \Lambda_{b_2}{}^e \overline{E}_{b_1,ae}   + e_a{}^{\mu} \partial_{\mu} \tilde{\Lambda}_{b_1 b_2}
\eqno(4.7)$$
In the process of carry  out this calculation one finds that the variation of ${\cal E}_{a,b_1 b_2}=0$ leads to a trivial dynamics unless  $\tilde{e}_1 = 1$ which is the value we now adopt.
\par
Setting the variation of ${\cal E}_{a,b_1 b_2} \dot =0$ we find the  gravity-dual gravity relation $E_{a,bc}\dot =0$, from which we started,  as well as  a dual graviton-  dual dual graviton duality relation which is given by
$$
\overline{E}_{a , b_1 b_2} \equiv  \overline{G}_{a,b_1 b_2} + \varepsilon_a{}^{e_1 e_2 e_3} G_{e_1 , e_2 e_3 , b_1 b_2} \dot = 0
\eqno(4.8) $$
\par
In the variation of equation (4.7) we also find the local  Lorentz   transformations
$$
e_a{}^{\mu} \partial_{\mu} \tilde{\Lambda}_{b_1 b_2} = - \varepsilon_{b_1 b_2}{}^{c_1 c_2} (  \Lambda_{c_2}{}^e G_{a,(c_1 e)}  -  G_{a , e_1 c_1, c_2 e_2} \Lambda^{e_1 e_2})  - 2 \Lambda_{e}{}_{[ b_1|} \overline{G}_{a,| b_2 ] e}.
\eqno(4.9)$$
As we have noted some of the equations we find only hold modulo certain local transformations and in the case of the gravity-dual gravity duality relation these include  local Lorentz transformations. The symbol $\dot =$ indicates that the equations only hold modulo the local transformations. 


\medskip
{\bf{5. The gravity and dual gravity equations of motion}}
\medskip
In this section we will use the symmetries of the non-linear realisation to find the equations of motion for the graviton and the dual graviton which are second order in derivatives. Since the level one local transformations with parameter  $\Lambda^{ab}$ change the level of the Cartan form on which it acts by plus or minus one,   the variation of the gravity equation must led  to the dual gravity equation. We begin with the
the usual Ricci tensor
$$
(\det e)R_a{}^b = (\det e) \{ e_a{}^{\mu} \partial_{\mu} (\omega_{\nu,}{}^{bd}) e_d{}^{\nu} - \partial_{\nu} (\omega_{\mu}{}^{bd}) e_d{}^{\nu} e_a{}^{\mu} + \omega_{a,}{}^b{}_c \omega_{d,}{}^{cd} - \omega_{d,}{}^b{}_c \omega_{a,}{}^{cd} \} \eqno(5.1)$$
In order to carry out its local $I_c(A_1^{+++})$  variation we must express the Ricci tensor  in terms of the Cartan forms of section three, the result is
$$
(\det e)R_a{}^b = (\det e)^{1 \over 2} e_a{}^{\mu} \partial_{\mu} [ (\det e)^{1 \over 2}  \omega_{d,}{}^{bd} ] - (\det e)^{1 \over 2}  e_d{}^{\nu} \partial_{\nu} [ (\det e)^{1 \over 2} \omega_{a,}{}^{bd}] $$
$$
+(\det e) \omega^{c,bd} \omega_{d,ca} +  G_{c,d}{}^{c} (\det e)^{1 \over 2} \omega_{a,}{}^{bd} - {1 \over 2}  G_{a,c}{}^c  (\det e)^{1 \over 2} \omega_{d,}{}^{bd} 
- {1 \over 2} G_{d,c}{}^c (\det e)^{1 \over 2} \omega_{a,}{}^{bd}  
\eqno(5.2)$$
where the expression for the spin connection in terms of the gravity Cartan forms is given in equation (4.2).
\par
We begin by considering the Ricci tensor as we expect  our equation of motion will turn out to be that this object  will vanish. As such we define 
$$
E_a{}^b\equiv (\det e)R_a{}^b 
\eqno(5.3)$$
and consider its variation under  local $I_c(A_1^{+++})$ transformations.  As we explained above in   carrying out the variation we must find the $l_1$ extension of the equations we are varying. We denote this $l_1$ extended object by ${\cal E}_a{}^b$. A help towards the result is achieved if one replaced the usual spin connection by its $l_1$ extension of equation (4.4) which has the variation of equation (4.5). After a somewhat lengthy calculation one finds that
$$
{\cal E}_{ab}^{\prime} \equiv (\det e) {\cal R}_{ab} = (\det e) \{ e_a{}^{\mu} \partial_{\mu} (\Omega_{\nu,}{}^{bd}) e_d{}^{\nu} - \partial_{\nu} (\Omega_{\mu}{}^{bd}) e_d{}^{\nu} e_a{}^{\mu} + \Omega_{a,}{}^b{}_c \Omega_{d,}{}^{cd} - \Omega_{d,}{}^b{}_c \Omega_{a,}{}^{cd} \} $$
$$
+ (\det e)^{1 \over 2} \hat{e}_a{}^{\nu} \partial_{\nu} \overline{G}_{[c,b]}{}^c  + \hat{G}_{a,}{}^c{}_d  \overline{G}_{[c,b]}{}^{d}  -  \hat{G}_{a,}{}^{d}{}_b \overline{G}_{[c,d]}{}^{c} + \hat{G}_{a,}{}^{d}{}_{c} \overline{G}_{[b,d]}{}^{c} - {1 \over 2} \hat{G}_{a,}{}^{c}{}_{c} \overline{G}_{[b,d]}{}^{d}  $$
$$
+ (\det e)^{1 \over 2} \hat{e}_b{}^{\nu }  \partial_{\nu}  \overline{G}_{[c,a]}{}^{c} + \hat{G}_{b,}{}^{c}{}_d \overline{G}_{[c,a]}{}^{d} - \hat{G}_{b,}{}^{d}{}_{a} \overline{G}_{[c,d]}{}^{c}  +  \hat{G}_{b,}{}^{d}{}_{c}  \overline{G}_{[a,d]}{}^{c} - {1 \over 2} \hat{G}_{b,}{}^{c}{}_c \overline{G}_{[a,d]}{}^{d} $$
$$
- \eta_{ab} ( (\det e)^{1 \over 2}  \hat{e}^{e}{}_{\nu}  \partial_{\nu}  \overline{G}_{[c,e]}{}^{c} +  \hat{G}^{e,}{}^{c}{}_{d}  \overline{G}_{[c,e]}{}^{d} - \hat{G}^{e,}{}^{d}{}_{e} \overline{G}_{[c,d]}{}^{c} + \hat{G}^{e,}{}^{d}{}_{c}  \overline{G}_{[e,d]}{}^{c} - {1 \over 2} \hat{G}^{e,c}{}_c \overline{G}_{[e,d]}{}^{d}  ) $$
$$
- ( \hat{G}^{e,}{}^{c}{}_{a} \overline{G}_{[b,c]e}  + \hat{G}^{e,}{}^{c}{}_{b} \overline{G}_{[a,c]e} + \eta_{ab} \hat{G}^{e,}{}_{[f_1f_2]} \overline{G}^{[f_1,f_2]}{}_{e} ) +  \hat{\partial}^e \overline{G}_{[a,b]e}  +  \hat{G}^{e,}{}_{e}{}^c \overline{G}_{[a,b]c}     $$
$$
+ {1 \over 2} \hat{G}^{c,}{}_{e}{}^e \overline{G}_{[a,b]c}  - {1 \over 2}   (\det e)^{1 \over 2} \Omega_{d,b}{}^{d} \hat{\overline{G}}^{e,}{}_{a e} - {1 \over 2} (\det e)^{1 \over 2} \Omega_{a,b}{}^{d} \hat{\overline{G}}^{e,}{}_{ed}
 \eqno(5.4) $$
has the variation
$$
\delta {\cal E}_{ab} ^{\prime} = -  4 \Lambda_e{}_a \overline{E}^{\prime} {}_b{}^e - 4 \Lambda_e{}_b \overline{E}^{\prime} {}_a{}^e + 4 \eta_{a b} \Lambda^{e_1}{}_{e_2} \overline{E}^{\prime} {}_{e_1}{}^{e_2} $$
$$
+ \Lambda^{e_1 e_2} \varepsilon_{bc}{}^{f_1f_2} ( E_{e_2,}{}_a{}^e (\det e)^{1 \over 2} \omega_{e_1,f_1f_2}  -  E_{e_1,f_1f_2} (\det e)^{1 \over 2} \omega_{e_2,}{}_a{}^c ) $$
$$
+ \Lambda^{e_1 e_2} \varepsilon_{ac}{}^{f_1f_2} ( E_{e_2,}{}_b{}^e (\det e)^{1 \over 2} \omega_{e_1,f_1f_2}  -  E_{e_1,f_1f_2} (\det e)^{1 \over 2} \omega_{e_2,}{}_b{}^c ) 
\eqno(5.5) $$
where we defined 
$$
\overline{E}^{\prime} {}_{a}{}^{b} \equiv (\det e)^{1 \over 2} e^{\nu [c}  \partial_{\nu}  \overline{G}_{[c,a]}{}^{b]} +  G^{[c,b]}{}_{d}  \overline{G}_{[c,a]}{}^{d}  -  G^{[c|,d}{}_{c}  \overline{G}_{[d,a]}{}^{|b]}  $$
$$
-  G^{[d|,c}{}_{a} \overline{G}_{[d,c]}{}^{|b]} +  {1 \over 2} G^{[d|}{}_{,c}{}^c \overline{G}_{[d,a]}{}^{|b]}. 
\eqno(5.6)$$
Converted to  world volume indices $\overline{E}^{\prime} {}_{a}{}^{b} $ takes the form 
$$
\overline{E}^{\prime} {}^\mu{}_{\tau} = \partial_{[\nu} ( (\det e)^{{1 \over 2}} \overline{G}^{[\nu,\mu]}{}_{\tau]} ). 
\eqno(5.7)$$
\par
We observe that the variation of ${\cal E}^{\prime} {}_{ab}$ contains our previously discussed  the first order gravity-dual gravity relation $E_{a,}{}_b{}^c $,  found in the previous section, as well as the  new object $\overline{E}^{\prime} {}_{a}{}^{b}$. We note that these occur in a different ways in relation to the parameter $ \Lambda_{ab}$.  As a result,   we may  take the equations  of motion to be 
${ E}_{ab} =0$ and $ \overline{E}^{\prime} {}_{a}{}^{b} =0$ as these are an invariant set of equations up to the level computed. 
The first equation is just Einstein's equation for gravity, as one might be expect, while the second equation would be that for the dual graviton.This conclusion would however, be premature. It over looks the fact that the $l_1$ extension of the Einstein equation could contain terms  $\hat G_{b, \bullet}X$ where $X$ is any function of the Cartan forms with derivatives that are with respect to the usual spacetime coordinates. These terms would lead in the variation of the Einstein equation  ${\cal E}_{ab} ^{\prime}$  to terms of the form $\Lambda ^{be} \overline G_e{}_{,\bullet }X$ where $\bullet$ is a $E_{11}$ index. Looking at the variation of equation (5.5) we see that such a term would result in an addition to the dual graviton equation $\overline E_a{}_b $ of a term of the form  index  $G_b{}_{, \bullet} X$.  We note this is a term which contains a spacetime derivative with an index that corresponds to the second index on   $\overline{E}^{\prime}_{a}{}^{b}$. The primes on  ${\cal {E}}^{\prime}_{a}{}^{b}$ and $ \overline{E}^{\prime}_{a}{}^{b}$ are to indicate that we have not so far taken account of this possibility and so these objects are not the final results. We will now take account of this possibility and find which terms can be added in the way suggested. 
\par
The  dual graviton equation is by definition the equation of motion for dual graviton and as a result it should have the same symmetries as the dual gravity field, that is, it should be symmetric in its two indices. While the effect of exchanging the  $a$ and $b$ indices is obvious for the $G \bar G $ terms in equation (5.6)  it is not so obvious for the first term. To clarify this  we rewrite the first term in equation (5.6) as 
$$
(\det e)^{{1 \over 2}} e^{\nu [c}  \partial_{\nu}  \overline{G}_{[c,a]}{}^{b]} = {1 \over 4} (\det e)^{{1 \over 2}}   e^{\nu c}  \partial_{\nu} \overline{G}_{c,a}{}^{b} 
-  {1 \over 4} (\det e)^{1 \over 2}  ( e^{\nu c}  \partial_{\nu}  \overline{G}_{a,c}{}^{b} + e^{\nu c}  \partial_{\nu}  \overline{G}^{b,}{}_{ac})
$$
$$
+ {1 \over 8} (\det e)^{1 \over 2} ( e^{\nu b}  \partial_{\nu}  \overline{G}_{a,c}{}^{c} + e_a{}^{\nu}  \partial_{\nu}  \overline{G}^{b,c}{}_{c}) 
$$
$$
- {1 \over 4} (\det e)^{1 \over 2}  (  e^{\nu b}  \partial_{\nu}  \overline{G}_{c,a}{}^{c} - e^{\nu}{}_{c}  \partial_{\nu}  \overline{G}^{b,}{}_{a}{}^c ) 
+ {1 \over 8} (\det e)^{1 \over 2} ( e^{\nu b}  \partial_{\nu}  \overline{G}_{a,c}{}^{c} - e_a{}^{\nu}  \partial_{\nu}  \overline{G}^{b,c}{}_{c}) 
\eqno(5.8)$$
The first three terms are  obviously symmetric under the interchange of  $a$ and $b$. While  
the effect of this interchange on the   last two terms   is not so clear we can further rewrite them using the Maurer Cartan equations.  
\par
The form  ${\cal V} = g^{-1} d g$ it obeys the Maurer Cartan equation $d {\cal V} = - {\cal V} \wedge {\cal V}$, or equivalently 
$\partial_{\mu} {\cal V}_{\nu} - \partial_{\nu} {\cal V}_{\mu} + {\cal V}_{\mu} {\cal V}_{\nu} - {\cal V}_{\nu} {\cal V}_{\mu} = 0 $. Using the form of $ {\cal V}$ of equation (3.12) we find, amongst other equations,  that 
$$
(\det e)^{{1 \over 2}} e_c{}^{\mu} \partial_{\mu} \overline{G}_{d,ab} - (\det e)^{{1 \over 2}} e_d{}^{\mu} \partial_{\mu} \overline{G}_{c,ab} + G_{c,d}{}^e \overline{G}_{e,ab}- G_{d,c}{}^e \overline{G}_{e,ab} - {1 \over 2} G_{c,e}{}^e \overline{G}_{d,ab} $$
$$
+ {1 \over 2} G_{d,e}{}^e \overline{G}_{c,ab} + G_{c,a}{}^e \overline{G}_{d,eb} + G_{c,b}{}^e \overline{G}_{d,ae} - G_{d,a}{}^e \overline{G}_{c,eb} - G_{d,b}{}^e \overline{G}_{c,ae} = 0 
\eqno(5.9)$$
Using this last equation we can rewrite the last two terms of equation (5.8) as
$$
- {1 \over 4} (\det e)^{{1 \over 2}} [  e^{\nu b} \partial_{\nu} \overline{G}_{c,a}{}^{c} -  e^{\nu}{}_c \partial_{\nu} \overline{G}^{b,}{}_{a}{}^{c}] = + {1 \over 4}  ( G^{b,}{}_{c}{}^e \overline{G}_{e,a}{}^{c}- G_{c,}{}^{be} \overline{G}_{e,a}{}^{c} - {1 \over 2} G^{b,e}{}_e \overline{G}_{c,a}{}^{c} $$
$$
+ {1 \over 2} G_{c,e}{}^e \overline{G}^{b,}{}_{a}{}^{c} + G^{b,}{}_{a}{}^e \overline{G}_{c,e}{}^{c} + G^{b,}{}^{c}{}^e \overline{G}_{c,ae} - G_{c,a}{}^e \overline{G}^{b,}{}_{e}{}^{c} - G_{c,}{}^{ce} \overline{G}^{b,}{}_{ae} ) 
\eqno(5.10) $$
and
$$
{1 \over 8} (\det e)^{{1 \over 2}} ( e^{\nu b} \partial_{\nu} \overline{G}_{a,c}{}^c - e_a{}^{\nu} \partial_{\nu} \overline{G}^{b,c}{}_{c}) = - {1 \over 8} (G^{b,}{}_{a}{}^e \overline{G}_{e,c}{}^c - G_{a,}{}^{be} \overline{G}_{e,}{}^c{}_c - {1 \over 2} G^{b,e}{}_e \overline{G}_{a,c}{}^c $$
$$
+ {1 \over 2} G_{a,e}{}^e \overline{G}^{b,}{}^c{}_c  + G^{b,}{}_{c}{}^e \overline{G}_{a,}{}^c{}_{e} - G_{a,c}{}^e \overline{G}^{b,}{}_{e}{}^c - G_{a,c}{}^e \overline{G}^{b,}{}^{c}{}_{e}) 
\eqno(5.11)  $$
\par
Using equations (5.10) and (5.11) and explicitly writing out the anti-symmetrisations of the $G\bar G$ terms we find that the dual graviton expression  
$\bar E^{\prime}{}_{ab}$ of equation (5.6) can written as
$$
\overline{E}^{\prime}{}_a{}^b =   {1 \over 4}  (\det e)^{1 \over 2} ( e^{\nu c}  \partial_{\nu}  \overline{G}_{c,a}{}^{b}  -  e^{\nu c}  \partial_{\nu} \overline{G}_{a,c}{}^{b} - e^{\nu c}  \partial_{\nu}  \overline{G}^{b,}{}_{ac} + {1 \over 2}  e^{\nu b}  \partial_{\nu}  \overline{G}_{a,c}{}^{c} + {1 \over 2} e_a{}^{\nu}  \partial_{\nu}  \overline{G}^{b,c}{}_{c}) $$
$$
+ {1 \over 8} [ G^{c,b}{}_e ( 2 \overline{G}_{c,a}{}^e   - 2 \overline{G}^{e,}{}_{a}{}_{c} - 2 \overline{G}_{a,c}{}^e) +  G^{b,}{}_{c}{}^e (- 2\overline{G}_{a,}{}^c{}_{e}  + 2 \overline{G}_{e,a}{}^{c} +  2 \overline{G}^{c,}{}_{ae}  )  $$
$$
+ G^{e,c}{}_e (- 2 \overline{G}_{c,a}{}^b + 2 \overline{G}_{a,c}{}^b) + G^{d,c}{}_a (-2 \overline{G}_{d,c}{}^b + 2\overline{G}_{c,d}{}^b) + G^{b,c}{}_a (2 \overline{G}_{d,c}{}^d - 2 \overline{G}_{c,d}{}^d) $$
$$
+ G^{d,c}{}_c (\overline{G}_{d,a}{}^b - \overline{G}_{a,d}{}^b + \overline{G}^{b,}{}_{a}{}_{d}) + G^{b,c}{}_c (- \overline{G}_{d,a}{}^d + \overline{G}_{a,d}{}^d + {1 \over 2} \overline{G}_{a,d}{}^d - \overline{G}_{d,a}{}^{d} ) + G_{a,}{}^{be}\overline{G}_{e,c}{}^c $$
$$
+ G^{b,}{}_{a}{}^e (- \overline{G}_{e,c}{}^c + 2 \overline{G}_{c,e}{}^{c} ) + G_{a,c}{}^e (\overline{G}^{b,}{}_{e}{}^c + \overline{G}^{b,}{}^{c}{}_{e} ) + G_{c,a}{}^e (- 2 \overline{G}^{b,}{}_{e}{}^{c} ) $$
$$
+ G_{c,}{}^{ce}( - 2 \overline{G}^{b,}{}_{ae} ) + G_{a,e}{}^e (- {1 \over 2} \overline{G}^{b,c}{}_c ) ]  
\eqno(5.12)$$ 
\par
Clearly this expression for $\bar  E^{\prime}{}_{ab}$ of equation (5.12)  is not symmetric under $a\leftrightarrow b$ and so setting it to zero can not lead to the dual graviton equation. However,  we can exploit the above ambiguity to add terms to the $l_1$ extension of the Einstein expression and so to the dual graviton expression of equation (5.12). The terms  of equation (5.12) can be divided in to three types

\item{(a)}  terms which contain a $G_{b,\bullet}$ factor  ,

\item {(b)}  terms which contain a $G_{a,\bullet}$ factor, 

\item{(c)} the remaining terms. 

The type (a) terms can all be removed by adding terms to ${\cal E}^{\prime}{}_a{}^b$ as explained above. These terms occur in equation (5.12) as the number 3, 6, 8, 10 , 12, 13, 14 terms as well as the last expression in term 7. The type (b) terms by definition contain a    $G_{a,\bullet}$ factor and they occur in equation (5.12) as the terms number  2  (only last expression), 4 (only last expression), 7 (only middle expression) and term 9. These terms are given by 
$$
+ {1 \over 8} (- 2 G^{c,b}{}_e \overline{G}_{a,c}{}^e + 2 G^{e,c}{}_e \overline{G}_{a,c}{}^b  - G^{d,c}{}_c \overline{G}_{a,d}{}^b   + G_{a,}{}^{be}\overline{G}_{e,c}{}^c  ) 
\eqno(5.13)$$
For each of these terms we can swop the 
$a$ and $b$ indices and add the resulting term to the dual graviton equation as it contains a $G_{b,\bullet}$ factor. Put another way, we can in effect  symmetrise type (b) terms  by hand. The effect is that we add the terms 
$$
+ {1 \over 8} (- 2 G^{c,}{}_{a}{}_e \overline{G}^{b,}{}_{c}{}^e + 2 G^{e,c}{}_e \overline{G}^{b,}{}_{ca}  - G^{d,c}{}_c \overline{G}^{b,}{}_{da} + G^{b,}{}_{a}{}^{e}\overline{G}_{e,c}{}^c  ) 
\eqno(5.14)$$
to the dual graviton equation. 
\par
The terms of the  type (c) are given by 
$$
+ {1 \over 2} (  G^{c,b}{}^e \overline{G}_{[c,e]}{}_a   -   G^{c,d}{}_a \overline{G}_{[c,d]}{}^b ) - {1 \over 4}  G^{e,c}{}_e \overline{G}_{c,a}{}^b + {1 \over 8} G^{d,c}{}_c \overline{G}_{d,a}{}^b 
\eqno(5.15)$$
The last two terms are symmetric in $a\leftrightarrow b$ while the first two terms can be written as 
$$
+ {1 \over 2} ( G^{c,b}{}^d \overline{G}_{[c,d]}{}_a + G^{c,}{}_a{}^d \overline{G}_{[c,d]}{}^b) - {1 \over 4} \varepsilon^{cd e_1 e_2} \overline{G}_{[e_1,e_2]a}  \overline{G}_{[c,d]}{}^b  +  {1 \over 2} ( E_{a,}{}^{cd} -{1 \over 2} G_{a,}{}^{[c,d]} )\overline{G}_{[c,d]}{}^b  
\eqno(5.16)$$
The first and second  terms in this expression are  symmetric, while the third term is the gravity-dual gravity duality relation and the fourth term can be viewed as a modulo transformation to which this duality relation holds.  While the first two terms contribute to the dual gravity equation of motion, the last two terms can be reinterpreted as terms that explicitly occur in the variation of the $l_1$ extended gravity equation of motion ${\cal E} _a{}_b$, see equation (5.18) below. 
\par
After carrying out all the above steps we add the above  terms to $\bar E^{\prime}{}_a{}^b$ to  find that the dual graviton equation which is given by 
$$
\overline{E}_a{}^b \equiv   {1 \over 4}  (\det e)^{1 \over 2} ( e^{\nu c}  \partial_{\nu}  \overline{G}_{c,a}{}^{b}  -  e^{\nu c}  \partial_{\nu} \overline{G}_{a,}{}^{b}{}_c - e^{\nu c}  \partial_{\nu}  \overline{G}^{b,}{}_{ac} + {1 \over 2}  e^{\nu b}  \partial_{\nu}  \overline{G}_{a,c}{}^{c} + {1 \over 2} e_a{}^{\nu}  \partial_{\nu}  \overline{G}^{b,c}{}_{c}) $$
$$
- {1 \over 4}  G^{e,c}{}_e \overline{G}_{c,a}{}^b + {1 \over 8} G^{d,c}{}_c \overline{G}_{d,a}{}^b  $$
$$
- {1 \over 4} \varepsilon^{cd e_1 e_2} \overline{G}_{[e_1,e_2]a}  \overline{G}_{[c,d]}{}^b + {1 \over 2} ( G^{c,b}{}^d \overline{G}_{[c,d]}{}_a + G^{c,}{}_a{}^d \overline{G}_{[c,d]}{}^b)  $$
$$
- {1 \over 4}(G^{c,}{}_{a}{}_e \overline{G}^{b,}{}_{c}{}^e + G^{c,b}{}_e \overline{G}_{a,c}{}^e) + {1 \over 4}  G^{e,c}{}_e ( \overline{G}^{b,}{}_{ca} +  \overline{G}_{a,c}{}^b)$$
$$
- {1 \over 8} G^{d,c}{}_c(\overline{G}^{b,}{}_{da} + \overline{G}_{a,d}{}^b  ) + {1 \over 8} ( G^{b,}{}_{a}{}^{e}  + G_{a,}{}^{be} ) \overline{G}_{e,c}{}^c  
=0
\eqno(5.17)$$
It is indeed symmetric under the interchange of $a$ and $b$. That one can use the ambiguity to find an expression that is symmetric under the interchange of $a$ and $b$ is very non-trivial. The corresponding $l_1$ extension of the Einstein equation,  denoted ${\cal E}_a{}^b$,  is given in appendix A. The variation of ${\cal E}_a{}^b$ obeys the equation 
$$
\delta {\cal E}_{ab} = -  4 \Lambda_e{}_a \overline{E}_b{}^e - 4 \Lambda_e{}_b \overline{E} {}_a{}^e + 4 \eta_{a b} \Lambda^{e_1}{}_{e_2} \overline{E} {}_{e_1}{}^{e_2} $$
$$
+ \Lambda^{e_1 e_2} \varepsilon_{bc}{}^{f_1f_2} ( E_{e_2,}{}_a{}^e (\det e)^{1 \over 2} \omega_{e_1,f_1f_2}  -  E_{e_1,f_1f_2} (\det e)^{1 \over 2} \omega_{e_2,}{}_a{}^c ) $$
$$
+ \Lambda^{e_1 e_2} \varepsilon_{ac}{}^{f_1f_2} ( E_{e_2,}{}_b{}^e (\det e)^{1 \over 2} \omega_{e_1,f_1f_2}  -  E_{e_1,f_1f_2} (\det e)^{1 \over 2} \omega_{e_2,}{}_b{}^c )  
$$
$$
-  2\Lambda_{ea} ( E_{b,}{}^{cd} -{1 \over 2} G_{b,}{}^{[c,d]} )\overline{G}_{[c,d]}{}^e-
  2\Lambda_{eb} ( E_{a,}{}^{cd} -{1 \over 2} G_{a,}{}^{[c,d]} )\overline{G}_{[c,d]}{}^e+
$$
$$+2\eta_{ab} \Lambda^{e_1}{}_{e_2}  ( E_{e_1,}{}^{cd} -{1 \over 2} G_{e_1,}{}^{[c,d]} )\overline{G}_{[c,d]}{}^{e_2}
\eqno(5.18) $$
It is equation (5.5) with the primes removed and an extra term involving the gravity-dual gravity duality relation. The equations $E_a{}^b=0$ and 
$\bar E_a{}^b=0$,  together with the gravity-dual gravity duality relation,  form a set of equations that are transformed into each other and we can take them to be our equations of motion. 
\par
The above process has one further ambiguity associated with terms that are both of type (a) and type (b), that is,  they are of the form $G_{a,\bullet} G_{b,\bullet }$. Clearly one can either remove them or symmetrise them. The net effect is that we can add the terms to the dual graviton equation that are of the form 
$$
+ c_1 (G_{b,c}{}^e \overline{G}_{a,ce} + G_{a,c}{}^e \overline{G}_{b,ce})
\eqno(5.19)$$
$$
+ c_2 (G_{b,c}{}^c \overline{G}_{a,d}{}^d + G_{a,c}{}^c \overline{G}_{b,d}{}^d)
\eqno(5.20)$$
where $c_1$ and $c_2$ are constants. 
\par
One very stringent check of the above dual graviton equation  (5.17) is that it is Lorentz invariant. Under the transformations
$$
\delta \bar G_{a,bc}= \Lambda _a{}^e \bar G_{e,bc}+\Lambda _b{}^e \bar G_{a,ec}+\Lambda _c{}^e \bar G_{a,be} , 
$$
$$ 
\delta G_{a,bc}= \Lambda _a{}^e \bar G_{e,bc}+\Lambda _b{}^e \bar G_{a,ec}+\Lambda _c{}^e \bar G_{a,be} + e_a{}^\mu \partial_\mu \Lambda^{cb}
\eqno(5.21)$$
One does not need the possible additional terms of equations (5.19) and (5.20). The former expression is not invariant and so we can not add it to the dual graviton equation, however, the latter terms is invariant and so it is still a possible addition. It would be interesting to discuss the other expected symmetries of the dual graviton equation of motion. Of particular interest are diffeomorphism and the gauge symmetry. The corresponding transformations are discussed in section seven. As we observe there the transformation of the dual graviton is not as a general relativity tensor. 
We hope to return to this point in a future publication. 
\par
It is instructive to express the dual graviton equation in terms of objects carrying world indices. Defining $\overline{F}_{\mu, \nu_1 \nu_2} = \partial_{\mu} A_{\nu_1 \nu_2}$ the dual graviton equation (5.17) in world indices reads as 
$$
\overline{E}_{\mu}{}_{\nu} =  g^{\rho \sigma}  \partial_{[\sigma|} \overline{F}_{[\rho,\nu]| \mu]}  + {1 \over 4} g^{\rho \sigma}  G_{\tau,}{}_{\rho}{}^{\tau} (  \overline{G}_{\nu,}{}_{\mu `\sigma} +  \overline{G}_{\mu,\sigma}{}_{\nu} - \overline{G}_{\sigma,\mu}{}_{\nu}) $$
$$
+ {1 \over 4} g^{\rho \sigma}  G_{\rho,\tau}{}^{\tau} ( - \overline{G}_{\nu,}{}_{\mu}{}_{\sigma} - \overline{G}_{\mu,}{}_{\nu}{}_{\sigma}   + \overline{G}_{\sigma,}{}_{\mu}{}_{\nu} )   +  {1 \over 4} g^{\rho \sigma}  G_{\rho,}{}_{\sigma}{}^{\tau} ( - \overline{G}_{\tau,\mu}{}_{\nu} + \overline{G}_{\mu,}{}_{\nu}{}_{\tau} + \overline{G}_{\nu,}{}_{\mu}{}_{\tau} ) $$
$$
- {1 \over 4} g^{\rho \sigma}  G_{\nu,}{}_{\rho}{}^{\tau} \overline{G}_{\mu,\tau}{}_{\sigma} - g^{\rho \sigma}  {1 \over 4} G_{\mu,}{}_{\rho}{}^{\tau}  \overline{G}_{\nu,}{}_{\tau}{}_{\sigma} + {1 \over 16}  g^{\rho \sigma}  G_{\nu,}{}_{\tau}{}^{\tau}  \overline{G}_{\mu,\rho}{}_{\sigma} + {1 \over 16} g^{\rho \sigma}  G_{\mu,}{}_{\tau}{}^{\tau} \overline{G}_{\nu,}{}_{\rho}{}_{\sigma} 
$$
$$
- {1 \over 4}  (\det e)^{-1} \varepsilon^{\tau_1 \tau_2 \tau_3 \tau_4} \overline{G}_{[\tau_1,\tau_2]\mu} \overline{G}_{[\tau_3,\tau_4]}{}_{\nu} 
\eqno(5.22)$$
\par
In reference [8] the equations for the dual graviton in eleven dimensions was discussed. In particular the gravity-dual gravity duality relation   and the dual graviton equation of motion are  derived. While the former duality relation is directly derived from $E_{11}$ transformations and is correct, the derivation of the latter equation of motion relies on some additional  steps that have not been used in other $E_{11}$ papers. In particular it relied on diffeomorphism symmetry and the non-linear form of certain modulo transformations. As is apparent  from this paper the result for the dual graviton equation in reference [8]  is likely to be  incorrect as these additional steps were not correctly applied. However, it should be straightforward to apply  the techniques used in this paper to derive the dual graviton equation of motion  in eleven dimensions. 
\par
Another way to find the dual gravity equation is to carry out its variation under the non-linear symmetries. However, this is a very complicated task. We illustrate how it goes in the next section at the linearised level.


\medskip
{\bf{6. Derivation of the Linearised Equations of Motion and Variations}}
\medskip
In this section we will carry out  the variation of the  dual gravity equation of motion under the local $I_c(A_1^{+++})$ transformations, but only at the linearised level. The dual graviton  transforms under the $I_c(A_1^{+++})$ transformations of equation (3.18) into terms involving the graviton and dual dual-graviton, and so the resulting variation may be expected to involve the second order in derivatives gravity and dual gravity equations as well as derivatives of the previously derived first order duality relations. At the linearised level the dual graviton equation is given by 
$$
\overline E_a{}^b {}_{(lin)}\equiv   \partial ^{[c} G_{[c, a] , }{}^{ b ]}=0
\eqno(6.1)$$
\par
The $l_1$ extension of the   linearised dual graviton equation, ${\cal E}_a{}^b{}_{(lin)}$  transforms  under the  $I_c(A_1^{+++})$  transformation $\Lambda^{ab}$ as 
$$
\delta \overline{{\cal E}}_a{}^b = {1 \over 2} \Lambda_{a}{}^{c}  R_{c}{}^{b}  + {1 \over 2}  R_{a}{}^{c} \Lambda^{b}{}_{c} + 3 \partial^{b} E_{dac_1,}{}^d{}_{c_2} \Lambda^{c_1 c_2} +  \partial^{b} E_{c_2,ac_1} \Lambda^{c_1 c_2}  $$
$$
- {3 \over 2} \partial^{d} (E_{dac_1,bc_2}    + E_{dbc_1,ac_2}  )  \Lambda^{c_1 c_2} +  {3 \over 2} \partial^{d} E_{abc_1,dc_2} \Lambda^{c_1 c_2} $$
$$
- {1 \over 4} \varepsilon_{abc_1}{}^e  \partial^{d} \overline{E}_{d,e}{}^{c_2} \Lambda^{c_1}{}_{c_2} +  \varepsilon_{abc_1}{}^e \overline{E}_e{}^{c_2} \Lambda^{c_1}{}_{c_2}  
\eqno(6.2) $$
where 
$$
\overline{{\cal E}}_a{}^b{}_{(lin)} = \overline{E}_a{}^b{}_{(lin)} + {1 \over 2} ( \hat{\partial}_a (G^s)^{[d,}{}_{d}{}^{b]}  + \hat{\partial}^{b}  (G^s)_{[d,}{}_{a]}{}^{d}    )  + {1 \over 4} \hat{\partial}^c ( G^{b,}{}_{(a}{}_{c)}  - (G^s)_{a,}{}^{b}{}_{c}   + 2 G^{b,}{}_{[ac]}  ) $$
$$
+ {1 \over 4} \hat{\partial}^{c} ( \varepsilon_{abc}{}^e   \overline{G}_{[e,d]}{}^{d} - 2  G^{[d,}{}_{da,}{}^{b]}{}_{c}  - 2 G^{d,}{}_{[d}{}^{b,}{}_{a]c} -  G^{b,}{}_{a}{}_{d,}{}^{d}{}_{c} ) 
\eqno(6.3)$$
 In these equation we define $G^s_{a,bc} = G_{a,(bc)}$ and
$$
E_{a_1 a_2 a_3,bc} = {1 \over 3!}  \varepsilon_{a_1 a_2 a_3}{}^e \overline{E}_{e,bc}. 
\eqno(6.4)$$ 
where $\overline{E}_{e,bc}$ is defined in  equation (4.8).  In the equations in this section all Cartan forms and duality relations are to be taken to only contain their linearised expressions. 
Setting $ \overline{{\cal E}}_a{}^b =0$ we find, at the linearised level,  the gravity equation of motion $E_a{}^b= R_a{}^b=0$ and the dual gravity- dual  dual gravity duality relation $E_{a, b_1b_2}\dot =0$ which can also be written in the form of equation (6.4). Thus the equations shows that the variation of the dual gravity equation of motion transforms into quantities that we already know to vanish. 
\par
We have carried out some parts of the above calculation at the non-linear level. It is a much more difficult calculation. One of the most difficult aspects is that one must take account of the transformations that the duality relations hold subject to. We note in particular that the dual dual graviton equation of motion involves three derivatives and the duality relations this field satisfies even with two derivatives will only hold modulo certain transformations. 


\medskip
{\bf{7. Gauge transformations for $A_1^{+++}$}}
\medskip
The equations of motion that resulted from the non-linear realisation of $E_{11}\otimes_s l_1$ were essentially uniquely determined by its symmetries which were given in equation (3.7). These transformations do not include the usual gauge transformations. However, the resulting equations when restricted to contain just derivatives with respect to the usual coordinates of spacetime are invariant under all the usual gauge symmetries, that is, diffeomorphisms and the standard gauge transformations of the form fields.
It was proposed in  [21] that the theory was invariant under a set of gauge transformations whose parameters were in a one to one correspondence with the $l_1$ representation. Indeed they were contained in the parameter $\Lambda^A$ and the transformation of the fields of the theory could be given in terms of the variation of the vierbein in the formula [16]
$$
E^{-1}{}_A{}^{\Pi} \delta E_{\Pi}{}^B = (D^{\underline{\alpha}})_A{}^B C_{\underline{\alpha} \underline{\beta}} (D^{\underline{\beta}})_C{}^D D_D \Lambda^C 
\eqno(7.1)$$
where $\Lambda^A = \Lambda^{\Pi}E_{\Pi}{}^A$, $C_{\underline{\alpha} \underline{\beta}}$ is the Cartan-Killing metric of $E_{11}$,   $D_A $ is a suitable covariant derivative and the $D^{\underline{\alpha}}$ are the $l_1$  representation matrices which appear in the $E_{11}\otimes_s l_1$ algebra in the commutator
$$
[R^{\underline{\alpha}},l_A] = - (D^{\underline{\alpha}})_A{}^B l_B
\eqno(7.2)$$
This formula does indeed lead to the usual diffeomorphisms and form gauge transformations. These proposed gauge transformations have not played a central role in the construction of the eleven [4], seven [22] and five  dimensional  [3] theories from the non-linear realisation. However,
it is expected that they will play a more   important role when a more systematic  construction of the dynamics is given at all levels. These first order in derivatives relations only hold modulo certain transformations and these are very closely linked to the above local transformations.
\par
In this section we wish to find the analogous gauge transformations for the non-linear realisation $A_1^{+++} \otimes_s l_1$. Indeed we can simply apply the same formula as for the $E_{11}\otimes_s l_1$ case but now for the Kac-Moody algebra $A_1^{+++} \otimes_s l_1$. The gauge parameter
$ \Lambda^C  $ contains the components
$$
\xi^a, \ \hat{\xi}_a,  \ \Lambda_{a_1 a_2, a_3}, \  \Lambda_{a_1 a_2 a_3}, \  \ldots
\eqno(7.3)$$
We expect the first component to be the parameter  for the usual diffeomorphisms and the second component to be the corresponding analogue for the dual graviton.
\par
At low levels the first few $D^{\underline{\alpha}}$ matrices in the $A_1^{+++} \otimes_s l_1$. algebra are found to be given by
$$
(D^a{}_b) = \left(
\matrix{
\delta^a{}_c \delta^b{}_d - {1 \over 2} \delta^a{}_b \delta^d{}_c & 0 \cr
0 & - \delta^c{}_b \delta^a{}_d - {1 \over 2} \delta^a{}_b \delta^c{}_d \cr}
\right),
$$
$$
\ \ (D^{ab}) =
\left(
\matrix{
0 & - \delta^{(a}{}_c \delta^{b)}{}_d \cr
0 & 0 \cr}\right), \ \ (D_{ab}) = \left(
\matrix{0 & 0 \cr
- 2\delta^{c}{}_{(a} \delta^{d}{}_{b)} & 0 \cr}
\right), \ \ \ldots \eqno(7.4)$$
\par
The Cartan-Killing metric of $A_1^{+++}$  can be found, as usual, by noting that it is  invariant under $E_{11}$ transformations and so obeys the equation
$$
g([R^{\underline{\alpha}},R^{\underline{\beta}}], R^{\underline{\gamma}}) = g(R^{\underline{\alpha}},[R^{\underline{\beta}}, R^{\underline{\gamma}}])
\eqno(7.5)$$
It is straight forward using the level zero $GL(4)$ and level one invariances  to prove that it must, at low levels,  take the form
$$
g_{\underline{\alpha},\underline{\beta}} = 
\left(
\matrix{
\delta^c{}_b \delta^a{}_d - {1 \over 2} \delta^a{}_b \delta^c{}_d & 0 & 0  \cr
0 & 0 & 1 \cr
0 & 1 & 0 \cr}
\right).
\eqno(7.6)$$
\par
Using the above facts we find that equation (7.1) implies that
$$
E^{-1}{}_a{}^{\Pi} \delta E_{\Pi}{}^b = D_a \xi^b -{1\over 2}\delta_a^b  D_e\xi^e - \tilde D^b\tilde \xi_a-{1\over 2} \delta_a^b \tilde D^e\tilde \xi_e
$$
$$
E^{-1}{}_a{}^{\Pi} \delta E_{\Pi}{}_{\tilde b}= 2D_{(a}\tilde  \xi_{b)} , \ \ E^{-1}{}^{\tilde a}{}^{\Pi} \delta E_{\Pi}{}^ b= 2\tilde D^{(a} \xi^{b)}
\eqno(7.7)$$
Evaluating the left hand sides of these equations using the vielbein of equation (3.14) we find that
$$
E^{-1}{}_a{}^{\Pi} \delta E_{\Pi}{}^b =e_a{}^{\mu} \delta e_{\mu}{}^b -{1\over 2} \delta_a^b e_c{}^{\mu} \delta e_{\mu}{}^c ,\
E^{-1}{}_a{}^{\Pi} \delta E_{\Pi}{}_{\tilde b}= -e_a{}^\mu e_b{}^\nu \delta A_{\mu\nu} ,\ E^{-1}{}^{\tilde a}{}^{\Pi} \delta E_{\Pi}{}^ b=0
\eqno(7.8)$$
The appearance of the zero in the last equation is a consequence of the fact that we used the  local $I_c(E_{11})$ transformation of the non-linear realisation to set to zero all the negative level fields in the group element. It is this step that  resulted in the upper triangular form of the vielbein as well as  its inverse as shown in equation (3.14). and so the above zero.
\par
We observe that  our gauge transformations of equation (7.1) does  not preserved our  gauge choice using the   local $I_c(A_1^{+++})$ transformations. The solution is to carry out a simultaneous correctional local $I_c(A_1^{+++})$ transformation so as to preserve the gauge choice. The local $I_c(A_1^{+++})$ transformations with parameter  $\Lambda^{\underline \alpha}$ act as
$$
E^{-1}{}_A{}^{\Pi} \delta E_{\Pi}{}^B = (D^{\underline \alpha} - D^{-\underline \alpha}) \Lambda^{\underline \alpha}
\eqno(7.9)$$
Taking the local transformation constructed from plus and minus one generators this transformation takes the form
$$
E^{-1}{}_A{}^{\Pi} \delta E_{\Pi}{}^B = (D^{ab} -D_{ab} )_A{}^B \Lambda _{ab}
\eqno(7.10 )$$
and we find that
$$
E^{-1}{}_a{}^{\Pi} \delta E_{\Pi}{}^b =0,\
E^{-1}{}_a{}^{\Pi} \delta E_{\Pi}{}_{\tilde b}= -\Lambda_{a\tilde b}  ,\ E^{-1}{}^{\tilde a}{}^{\Pi} \delta E_{\Pi}{}^ b=2\Lambda^{\tilde ab}, \
E^{-1}{}^{\tilde a}{}^{\Pi} \delta E_{\Pi}{}_{\tilde b} =0,
\eqno(7.11)$$
\par
Carrying out a  simultaneous $I_c(A_1^{+++})$ with the parameter $\Lambda^{ab}= -\tilde D^{(a}\Lambda^{b)}$ we find that
$E^{-1}{}^{\tilde a}{}^{\Pi} \delta E_{\Pi}{}^ b=0$,  as it should,  and that the fields transform as
$$
e^{-1}{}_a{}^{\mu} \delta e_{\mu}{}^b = D_a \xi^b - \tilde {D}^b \tilde {\xi}_a - \delta_a{}^b \tilde {D}^c \hat{\xi}_c, \ \ \ \ 
\delta A_{\mu \nu} = e_{\mu}{}^a e_{\nu}{}^b (\tilde {D}_{(a} \xi_{b)} - 2 D_{(a} \tilde {\xi}_{b)}).
\eqno(7.12) $$
These transformations are not completely defined as we have not specified what are the connections the covariant derivatives utilise. We notice that the transformation of the dual graviton field  is not that of  a rank two symmetric  tensor under general relativity. In particular it does not contain terms of the form of an ordinary spacetime derivative acting on the parameter $\xi$.   This is to be expected as the dual graviton gives an alternative description of gravity rather than a field which couples in the normal way to gravity. 
\par
In reference  [21] the linearised gauge transformations of the fields of the non-linear realisation of $E_{11}\otimes_s l_1$ were derived starting  from the fact that they are  constructed from the
 derivatives $\partial_{\Pi}$ acting on the  parameters $\Lambda^{\Sigma}$.  The coefficients in front of such terms were fixed by demanding that if these quantities  transform according to the $\bar l_1$ representation and $l_1$ representation respectively then the transformations must transform in the adjoint representation of $E_{11}$, that is, like the fields themselves. The results agree with the transformations of equation (7.12) when linearised. The reader who wants to check this ascertion may find the transformations of the coordinates  under  the rigid $A_1^{+++}$ transformation $g_0= e^{a^{a_1 a_2} R_{a_1 a_2}}$ useful; they are given by   
$$
\delta x^a = 2 \tilde {x}_c a^{ca}, \ \ \delta \tilde {x}_a = 0, \ \ \ldots
\eqno(7.13)$$
This leads to the following transformations of the  derivatives transform
$$
\delta (\partial_a) = 0, \ \ \delta (\tilde {\partial}^a ) = - 2 a^{ac} \partial_c , \ \ \ldots
\eqno(7.14)$$
The reader can follow the procedure given in reference [21].

\medskip
{\bf 8. An alternative approach to the dual graviton}
\medskip
Rather than start form an $E_{11}$ view point we will now present a theory that contains the fields of gravity and dual gravity and has some of the expected symmetries. This is different to that of the non-linear realisation in that does not involve any extension of spacetime and the symmetries are proposed in an ad hoc way rather than being part of a deeper structure.
It will be instructive to  first recall some  very well known facts about  gravity. This is described by a vierbein $e_{\mu}{}^a$ which can be used to define a spin connection and curvature according to the equations
$$
D_{[\mu} e_{\nu]}{}^a \equiv \partial_{[\mu} e_{\nu]}{}^a + \omega_{[\mu|}{}^a{}_{b} e_{|\nu]}{}^b = 0, \quad \quad D_{[\mu} \omega_{\nu]}{}^{a_1a_2} - {1 \over 2} R_{\mu \nu}{}^{a_1 a_2} = 0,
\eqno(8.1)$$
These equations are invariant under the local Lorentz  transformations given by
$$
\delta e_{\nu}{}^a = \Lambda^a{}_b e_{\nu}{}^b, \quad \quad \delta \omega_{\mu}{}^{a b} = - D_{\mu} \Lambda^{a b}
= - (\partial_\mu  \Lambda^{a b}+ \omega_\mu {}^a{}_c \Lambda ^{cb} +  \omega_\mu {}^b{}_c \Lambda ^{ac}), \ \
$$
$$
\delta R_{\mu\nu} {}^{ab}= -R_{\mu\nu}{}^{a c} \Lambda _{c}{}^{ b}- R_{\mu\nu}{}^{c b}  \Lambda {}_c{}^{a}
\eqno(8.2)$$
where  $\Lambda^a{}_b$ is the  parameter local of the Lorentz transformation. We recall that $[D_\mu , D_\nu ]T^a = R_{\mu\nu} {}^a {}_c T^c $ for any tensor $T^a$ with obvious generalisations for   tensors with more indices. 
\par
Equations (8.1)  are also invariant under the transformations
$$
\delta e_{\nu}{}^a = D_\mu \xi^a \equiv \partial_\mu \xi^a + \omega_\mu {}^a{}_b \xi^b , \ \
\delta \omega_{\mu}{}_{, a b}= R_{\mu c , ab}\xi^c
$$
$$
\delta R_{\mu\nu}{}^{ab}= \xi^c D_c  R_{\mu \nu}{}^{a b} - R_{\nu c }{}^{ab}D_\mu  \xi^c + R_{\mu c }{}^{ab}D_\nu  \xi^c
\eqno(8.3)$$
which are just a combination of a  usual diffeomorphism and a Local Lorentz rotation.
The invariant  field equations  are  given by
$$R_{\mu\nu}{}^{ab} e_b{}^{\nu} = 0
\eqno(8.4)$$.
\par
  We now introduce  the field  $\tilde{e}_{\mu}{}^a$ corresponding to  the dual graviton. We define a dual spin connection $\tilde{\omega}_{\mu}{}^a{}_b$ and dual curvature $ \tilde{R}_{\mu \nu}{}^{ab}$ as follows
$$
D_{[\mu} \tilde{e}_{\nu]}{}^a + \tilde{\omega}_{[\mu|}{}^a{}_b e_{|\nu]}{}^b = \partial_{[\mu} \tilde{e}_{\nu]}{}^a  + \omega_{[\mu|}{}^a{}_b \tilde{e}_{|\nu]}{}^b + \tilde{\omega}_{[\mu|}{}^a{}_b e_{|\nu]}{}^b= 0, \ \ \  D_{[\mu} \tilde{\omega}_{\nu]}{}^{ab} - {1 \over 2} \tilde{R}_{\mu \nu}{}^{ab} = 0,
\eqno(8.5)$$
where $D_{\mu}$ the usual covariant derivative of general relativity.
\par
These equations together with those of equation (8.1) are invariant under  the local  symmetries
$$
\delta \tilde{e}_{\mu}{}^a = \tilde{\Lambda}^a{}_b e_{\nu}{}^b, \quad \delta \tilde{\omega}_{\mu}{}^{ab} = - D_{\mu} \tilde{\Lambda}^{ab}= -(\partial_\mu \tilde{\Lambda}^{ab}+ \omega_\mu {}^a{}_c \tilde \Lambda ^{cb}+  \omega_\mu {}^b{}_c \tilde \Lambda ^{ac} ),
$$
$$
\delta \tilde R_{\mu\nu} {}^{ab} = -R_{\mu\nu } {}^{ac}\tilde \Lambda _{c}{}^b-R_{\mu\nu } {}^{cb}\tilde \Lambda _{c}{}^a
\eqno(8.6)$$
where  $\tilde{\Lambda}^{ab} $ is the parameter.  They are also invariant  under the transformations
$$
\delta \tilde{e}_{\mu}{}^a =D_\mu \tilde \xi^a , \ \ \delta \tilde{\omega}_{\mu}{}^{ab} = -R_{\mu c }{}^{ab} \tilde \xi ^c
$$
$$
\delta \tilde R_{\mu\nu} {}^{ab}=  +\tilde \xi^c D_c  R_{\mu \nu}{}^{a b} - R_{\nu c }{}^{ab}D_\mu  \tilde \xi^c + R_{\mu c }{}^{ab}D_\nu  \tilde \xi^c
\eqno(8.7)$$
\par
We take as our equation of motion
$$
\tilde E_\mu {}^a\equiv \tilde R_{\mu\nu} {}^{ab} e_b{}^\nu - R_{\mu\nu} {} ^{ab} e_d{}^\nu \tilde e_\tau {}^d e_b {}^\tau =0
\eqno(8.8)$$
One can verify that it is invariant under the usual diffeomorphism and local Lorentz transformations but it is also invariant under the
transformations of equations (8.6) and (8.7)
\par
We can solve the first of the equations in (7.5) for the  dual spin connection in much the way that one solves for the usual spin connection using  equation (8.1). One finds that
$$
\tilde{\omega}_{a,bc} = - {F}_{b,(ca)} + {F}_{c,(ba)} + {F}_{a,[bc]}
\eqno(8.9)$$
where $F_{a,b}{}^{c}= e_a{}^\mu e_b{}^\tau (\partial _\mu \tilde e_\tau {}^c+ \omega_{\mu ,}{}^c{}_ d\tilde e _\tau {}^d)=  e_a{}^\mu e_b{}^\tau D_\mu \tilde e_\tau {}^c $.
\par
If we define
$$
\tilde e _{\mu,\nu}= \tilde e_\mu {}^c e_\nu {}_c
\eqno(8.10)$$
then one can show that the dual spin connection can be written as
$$
\tilde \omega _{\mu , bc}= e_b{}^\kappa e_c{}^\lambda 2( - \partial_{[\kappa} \tilde e^S_{\lambda ] }{}_\mu +\partial_\mu \tilde e^A _{\kappa\lambda} + \Gamma _{\mu [\kappa}^\rho \tilde e_{\lambda ] , \rho})
\eqno(8.11)$$
where $\tilde e^S_{\kappa \lambda}= \tilde  e_{(\kappa \lambda )}$, $\tilde e^A_{\kappa \lambda}= \tilde  e_{[\kappa \lambda ]}$ and
$ \Gamma _{\mu \kappa}^\rho$ is the usual Christoffel connection which obeys the relation
$\partial_\mu e_\nu^a +\omega_{\mu , }{}^a{}_b e_\nu{}b = \Gamma _{\mu\nu}^\rho e_\rho{}^a$.
Substituting the above  expression for the dual spin connection   in the equation of motion of equation (8.8) one finds the equation of motion for the dual graviton as discussed in this section.
\par
It would be interesting to find the relationship between the formulation of dual gravity given in this section and the one that follows from the non-linear realisation of  $A_1^{+++}\otimes_s l_1$ and is the main subject of this paper.


\medskip
{\bf {9. Discussion}}
\medskip
In this paper we have carried out the non-linear realisation of the semi-direct product of the very extended Kac-Moody algebra $A_1^{+++}$ with its vector representation, denoted by $A_1^{+++}\otimes_s l_1$. We found that the resulting equations of motion at lowest level describe gravity when the derivatves with respect to the higher level coordinates are discarded. At the next level we found the fully non-linear equation of motion for the dual graviton.  We also find that the gravity and dual gravity fields satisfy a first order in derivative duality relation. The  fields  in the non-linear realisation  up to level four are  listed in equation (3.4). As is apparent from this list  the  fields at higher levels  have an increasing number of indices that obey more and more complicated symmetrisation and anti-symmetrisation conditions.  The third field listed in equation (3.4) is the dual dual graviton and at higher levels we find further duals of gravity. These  fields  have the form $A_{a_1a_2, \ldots ,d_1d_2, (e_1e_2)}$ and in the listing of equation (3.4) they are the second, third, fourth and sixth fields. The occurrence of such dual fields was observed in the context of $E_{11}$ [23] and for other non-linear realisations of the semi-direct products of extended algebras and their vector representations   in reference [24], although this reference did not include studies involving the very extended algebras,  $A_{D-3}^{+++}$,  associated with gravity.  It would be good to know what is the  physical meaning of the fourth field $A_{a_1a_2a_3, b_1b_2, c}$ in the listing of equation (3.4). 
\par
The spacetime coordinates belong to   the vector representation and are listed in equation (3.6). By construction these are in one to one correspondence with the generators in the vector representation. For the case of $E_{11}$ it is clear that the multiplet of brane charges belong to the vector representation of $E_{11}$. As a result, it is very likely that the brane charges of the non-linear realisation of $A_1^{+++}\otimes_s l_1$ also contains all brane charges of this theory and as such are given at low levels  in equation (2.10). The first entry is just the just momentum operator corresponding to translations in our usual spacetime and is associated with the gravity field. The next entry $Z_a$ is associated with the dual graviton and it is the charge carried by the Taub-Nut solution. It would be good to know what is the physical significance of the higher charges. 
\par
When we truncate to only the lowest level field, that is, that of gravity and retain only the usual coordinate of our usual four dimensional spacetime the non-linear realisation is just Einstein's theory of gravity. However, the full non-linear realisation contains an infinite number of  fields which depend on a spacetime that has an infinite number of coordinates. It is also invariant under the infinite algebra, namely  $A_1^{+++}\otimes_s l_1$. 
As such the full nonlinear realisation of $A_1^{+++}\otimes_s l_1$ contains much more than Einstein's theory of gravity. For example,  it contains an infinite number of new degrees of freedom corresponding to the infinite number of brane charges and their corresponding solutions. It has been realised that to explain features of gravity such as the entropy of black holes one needs a theory that goes beyond our usual understanding of Einstein's theory. It would be interesting to see if the additional content of  the nonlinear realisation of $A_1^{+++}\otimes_s l_1$ can be used in this way and in particular if it can be used to explain black hole entropy. The brane charges correspond to weights of $A_1^{+++}$ and as one requires large brane charges this means weights of high level. One might wonder if the calculation of the black hole entropy can be formulated as 
a combinatoric problem constructing such high level weights from the more fundamental weights and roots of the $A_1^{+++}$ algebra. 
\par
A very interesting discovery in the 1950's was the existence of asymptotic (BMS) charges in gravity. This work has been considerably extended in more recent times,  see reference  [25] and the references it contains. Very recently it has been shown that one should also include the  asymptotic charges associated with the Taub-Nut solution [26]. Could it be possible that the brane charges for the non-linear realisation of  $A_1^{+++}\otimes_s l_1$ studied in this paper are closely related and even the same as the asymptotic charges that are being studied. The asymptotic charges and the charges in the vector representation appear to  agree at the first two levels. 
\par
The non-linear realisations of $G^{+++}\otimes_s l_1$, where $G$ is any Lie algebra in the Cartan List, possess an invariant tangent space metric which was constructed at low levels in reference [27] for many of these non-linear realisations including for $A_1^{+++}\otimes_s l_1$. The tangent vectors transform under $I_c(A_1^{+++})$ and arise from  the $l_1$ representation. It we label their components by 
$$
T^a , \bar T_a , T_{a_1a_2a_2} ,  T_{a_1a_2 , b} , \ldots 
\eqno(9.1)$$
the invariant tangent space metric is given by 
$$
L^2\equiv T_a T^a+ 2\bar T_a \bar T^a+ 4T_{a_1a_2a_2}T^{a_1a_2a_2}+{16\over 3}  T_{a_1a_2 , b}  T_{a_1a_2 , b} +\ldots
\eqno(9.2)$$
This expression provides an invariant bilinear in the brane charges $l_\Pi$ which do transform in the vector representation  by taking $T_A=E_A{}^\Pi l_\Pi$ where $E_A{}^\Pi$ is the inverse vierbein of equation (3.14). 
\par
It was found in reference [28] that, for the case of $E_{11}$, setting $L^2=0$ coincided  at low levels with the half BPS conditions that can be derived from the supersymmetry algebra. Hence it is natural to take the condition $L^2=0$ to be the analogue of the half BPS conditions  for the case of the non-linear realisation of  $A_1^{+++}\otimes_s l_1$. This condition begins with the square of the momentum generators and the next term contains the square of the Taub-Nut charge. Such a condition has been proposed from the view point of gravitational duality relating these two charges [28,29]. As was explained in the last of these references this condition can not be derived from the usually supersymmetry algebra. It does however, follow naturally from the non-linear realisation studied in this paper. It has been know for a long time that the supersymmetry algebra does not contain all the required brane charges but that they are contained in E theory. The presence of such a relation involving the momenta and Taub-NUT charge is generic to the $G^{+++}\otimes_s l_1$ non-linear realisations including in E theory. Thus the Taub-Nut charge is just one of an infinite number of charges  that is missing from the supersymmetry algebra. It would also be interesting to find  for the $A_1^{+++}\otimes_s l_1$ non-linear realisation the analogue of equation (13) and the "quarter BPS" condition of equation (39) of reference [28] and also  interpret the former in the sense of reference [31].


\medskip
{\bf {Acknowledgments}}
\medskip
We wish to thank Paul Cook, Hadi and Mahdi  Godazgar and Dionysis Anninos for discussions. Peter West wishes to thank the SFTC for support from Consolidated grants number ST/J002798/1 and ST/P000258/1, while Keith Glennon would like to thank the STFC for support during his PhD studies.



\medskip
{\bf {Appendix A}}
\medskip
\par
In this appendix we will  list the terms which are referred to,   but not explicitly stated,  in section five when we were  constructing the symmetric dual graviton equation $\overline{E}_a{}^b$ of (5.17).  In particular we will discuss in more detail the construction of the $l_1$ extended Einstein equation ${\cal E}_{ab}$ of (5.18) which lead in its variation to the  terms  we added to  $E^{\prime}_a{}^b$ of (5.12) to find the dual graviton equation of motion   $\overline{E}_a{}^b$ .
\par 
We recall that equation  (5.12) contained terms that were derivatives of the dual graviton Cartan form and also  terms that were divided into the  types (a), (b) and (c). The terms of type (a) occur in equation (5.12) as the number  3, 6, 8, 10 , 12, 13, 14 terms as well as the last expression in term 7. We remove these terms by adding their negative to $\bar E^{\prime}_a{}^b$, that is, we add the terms 
$$
-{1 \over 4} G^{b,}{}_{c}{}^e (- \overline{G}_{a,}{}^c{}_{e}  +  \overline{G}_{e,a}{}^{c} +   \overline{G}^{c,}{}_{ae}  )  + {1 \over 4} G^{b,c}{}_a ( \overline{G}_{c,d}{}^d -  \overline{G}_{d,c}{}^d)  $$
$$
- {1 \over 8} G^{b,c}{}_c (-2 \overline{G}_{d,a}{}^d +  {3 \over 2} \overline{G}_{a,d}{}^d  ) $$
$$
- {1 \over 8} G^{b,}{}_{a}{}^e (- \overline{G}_{e,c}{}^c + 2 \overline{G}_{c,e}{}^{c} ) -{1 \over 4} G_{a,c}{}^e \overline{G}^{b,}{}_{e}{}^c   +{1 \over 4} G_{c,a}{}^e  \overline{G}^{b,}{}_{e}{}^{c} $$
$$
+ {1 \over 4} G_{c,}{}^{ce}  \overline{G}^{b,}{}_{ae} + {1 \over 16} G_{a,e}{}^e  \overline{G}^{b,c}{}_c  
- {1 \over 8} G^{d,c}{}_c \overline{G}^{b,}{}_{a}{}_{d}
\eqno(A.1)$$
\par 
The type (b) terms by definition contain a    $G_{a,\bullet}$ factor and they occur in equation (5.12) as the terms number  2  (only last expression), 4 (only last expression), 7 (only middle expression) and term 9,   we list them again here for convenience 
$$
+ {1 \over 8} (- 2 G^{c,b}{}_e \overline{G}_{a,c}{}^e + 2 G^{e,c}{}_e \overline{G}_{a,c}{}^b  - G^{d,c}{}_c \overline{G}_{a,d}{}^b   + G_{a,}{}^{be}\overline{G}_{e,c}{}^c  ).
\eqno(A.2)$$
As explained in equation (5.14) we can obtain an expression which is symmetric in $a$ and $b$ by adding to  $E^{\prime}_a{}^b$ the terms 
$$
+ {1 \over 8} (- 2 G^{c,}{}_{a}{}_e \overline{G}^{b,}{}_{c}{}^e + 2 G^{e,c}{}_e \overline{G}^{b,}{}_{ca}  - G^{d,c}{}_c \overline{G}^{b,}{}_{da} + G^{b,}{}_{a}{}^{e}\overline{G}_{e,c}{}^c  ) 
\eqno(A.3)$$
The net effect of this is that we find in $\bar E_a{}^b$ the $a,b$ symmetric terms 
$$
+ {1 \over 4}   (G^{c,b}{}_e \overline{G}_{a,c}{}^e + G^{c,}{}_{a}{}_e \overline{G}^{b,}{}_{c}{}^e) + {1 \over 4} ( G^{e,c}{}_e \overline{G}_{a,c}{}^b + G^{e,c}{}_e \overline{G}^{b,}{}_{ca} ) $$
$$
- {1 \over 8} ( G^{d,c}{}_c \overline{G}_{a,d}{}^b + G^{d,c}{}_c \overline{G}^{b,}{}_{da})  + {1 \over 8} ( G_{a,}{}^{be}\overline{G}_{e,c}{}^c  + G^{b,}{}_{a}{}^{e}\overline{G}_{e,c}{}^c ) 
\eqno(A.4) $$
\par 
The terms of type (c) have been listed in (5.15) and re-expressed in (5.16) and, as explained there, two  of these terms are symmetric under the interchange of $a$ and $b$ and are part of the dual graviton equation and the final terms can be viewed as part of the variation of the gravity equation of motion. 
\par
As noted in section five some of the terms are of both (a) and (b) type, listed as they appear in  $\overline{E}^{\prime}_a{}^b$ they are
$$
- {1 \over 4} G^{b,}{}_{c}{}^e \overline{G}_{a,}{}^c{}_{e} + {1 \over 4} G_{a,c}{}^e \overline{G}^{b,c}{}_{e} +{3 \over 16} G^{b,c}{}_c \overline{G}_{a,d}{}^d  - {1 \over 16} G_{a,c}{}^c \overline{G}^{b,d}{}_d  . 
\eqno(A.5) $$
Such terms can be treated as either type  (a) terms, that is in effect by removal, or as type (b) terms, that is, symmetrisation. The effect of this ambiguity is that the expression for $\bar E_a{}^b$ can contains two  terms, listed in equations (5.19)  and (5.20) whose coefficients are not fixed.  
The first of these terms was ruled out by considerations of Lorentz symmetry and so we only take account of the second term and as a result we add to $\bar E^{\prime}{}_a{}^b$ the term 
$$
 c_2 (G^{b,c}{}_c \overline{G}_{a,d}{}^d + G_{a,}{}^{c}{}_c \overline{G}^{b,}{}_{d}{}^d) 
\eqno(A.6) $$ 
\par
The result of all the above consideration is that we obtain the dual graviton equation of motion by  adding to  $\overline{E}^{\prime}{}_a{}^b$ the terms in equations (A.1), (A.3) and (A.6) as well as the term involving the gravity-dual gravity relation which arises in the type (c) terms discussed above. 
\par
The dual gravity equation of motion has been found by varying the gravity equation of motion and so the above  additions to the dual gravity equation of motion arise in the variation of the gravity equation of motion by adding $l_1$ terms to this equation, that is adding more such terms to the  ${\cal E}_{ab}^{\prime}$ of equation (5.4). The resulting $l_1$ extension of the gravity equation of motion is given, using equation (3.20),  to be

$$
{\cal E}_{ab} \equiv {\cal E}_{ab}^{\prime} +  {1 \over 2}  [ G^{c,}{}_{be}  \hat{\overline{G}}_{a,}{}_{c}{}^e - G^{e,c}{}_e \hat{\overline{G}}_{a,}{}_{cb} + {1 \over 2} G^{d,c}{}_c  \hat{\overline{G}}_{a,}{}_{db} - {1 \over 2} \hat{G}_{a,}{}_b{}^{e}\overline{G}_{e,c}{}^c $$
$$
+ \hat{G}_{a,}{}_{c}{}^e ( - \overline{G}_{b,}{}^c{}_e + \overline{G}_{e,b}{}^{c} +  \overline{G}^{c,}{}_{be}) + {1 \over 2} G^{d,c}{}_c  \hat{\overline{G}}_{a,}{}_{bd} - 2 \hat{G}_{a,}{}^{c}{}_b \overline{G}_{[c,d]}{}^d  $$
$$
-  G_{c,}{}^{ce} \hat{\overline{G}}_{a,}{}_{be} + {1 \over 2} \hat{G}_{a,}{}^{c}{}_c ( - 2 \overline{G}_{d,b}{}^d  + {3 \over 2} \overline{G}_{b,d}{}^d ) + {1 \over 2} \hat{G}_{a,}{}_{b}{}^e ( - \overline{G}_{e,c}{}^c + 2 \overline{G}_{c,e}{}^{c} ) $$
$$
-  G_{c,b}{}^e \hat{\overline{G}}_{a,}{}_{e}{}^{c}  + G_{b,c}{}^e \hat{\overline{G}}_{a,}{}_{e}{}^c - {1 \over 4} G_{b,e}{}^e  \hat{\overline{G}}_{a,}{}^{c}{}_c - 4 c_2 (\hat{G}_{a,}{}^{c}{}_c \overline{G}_{b,d}{}^d + G_{b,}{}^{c}{}_c \hat{\overline{G}}_{a,}{}_{d}{}^d) ] $$ 
$$
 +  {1 \over 2}  [ G^{c,}{}_{ae}  \hat{\overline{G}}_{b,}{}_{c}{}^e - G^{e,c}{}_e \hat{\overline{G}}_{b,}{}_{ca} + {1 \over 2} G^{d,c}{}_c  \hat{\overline{G}}_{b,}{}_{da} - {1 \over 2} \hat{G}_{b,}{}_a{}^{e}\overline{G}_{e,c}{}^c $$
$$
+ \hat{G}_{b,}{}_{c}{}^e ( - \overline{G}_{a,}{}^c{}_e + \overline{G}_{e,a}{}^{c} +  \overline{G}^{c,}{}_{ae}) + {1 \over 2} G^{d,c}{}_c  \hat{\overline{G}}_{b,}{}_{ad} - 2 \hat{G}_{b,}{}^{c}{}_a \overline{G}_{[c,d]}{}^d $$
$$
-  G_{c,}{}^{ce} \hat{\overline{G}}_{b,}{}_{ae} + {1 \over 2} \hat{G}_{b,}{}^{c}{}_c ( - 2 \overline{G}_{d,a}{}^d  + {3 \over 2} \overline{G}_{a,d}{}^d )  +  {1 \over 2} \hat{G}_{b,}{}_{a}{}^e ( - \overline{G}_{e,c}{}^c + 2 \overline{G}_{c,e}{}^{c} )  $$
$$
-  G_{c,a}{}^e \hat{\overline{G}}_{b,}{}_{e}{}^{c} + G_{a,c}{}^e \hat{\overline{G}}_{b,}{}_{e}{}^c - {1 \over 4} G_{a,e}{}^e  \hat{\overline{G}}_{b,}{}^{c}{}_c - 4 c_2 (\hat{G}_{b,}{}^{c}{}_c \overline{G}_{a,d}{}^d + G_{a,}{}^{c}{}_c \hat{\overline{G}}_{b,}{}_{d}{}^d) ] $$ 
$$
-  {1 \over 2} \eta_{ab} [ G^{c,}{}_{e_1 e_2}  \hat{\overline{G}}^{e_1,}{}_{c}{}^{e_2} - G^{e_1,c}{}_{e_1} \hat{\overline{G}}^{e_2,}{}_{ce_2} + {1 \over 2} G^{d,c}{}_c  \hat{\overline{G}}^{e,}{}_{de} - {1 \over 2} \hat{G}^{e_1,}{}_{e_1}{}^{e_2}\overline{G}_{e_2,c}{}^c $$
$$
+ \hat{G}^{e_1,}{}_{c}{}^{e_2} ( - \overline{G}_{e_1,}{}^c{}_{e_2} + \overline{G}_{e_2,e_1}{}^{c} +  \overline{G}^{c,}{}_{e_1e_2}) + {1 \over 2} G^{d,c}{}_c  \hat{\overline{G}}^{e,}{}_{ed} - 2 \hat{G}^{e,}{}^{c}{}_e \overline{G}_{[c,d]}{}^d  $$
$$
-  G_{c,}{}^{ce_1} \hat{\overline{G}}^{e_2,}{}_{e_2e_1} + {1 \over 2} \hat{G}^{e,c}{}_c ( - 2 \overline{G}_{d,e}{}^d  + {3 \over 2} \overline{G}_{e,d}{}^d ) +  {1 \over 2} \hat{G}^{e_1,}{}_{e_1}{}^{e_2} ( - \overline{G}_{e_2,c}{}^c + 2 \overline{G}_{c,e_2}{}^{c} ) $$
$$
-  G_{c,e_1}{}^{e_2} \hat{\overline{G}}^{e_1,}{}_{e_2}{}^{c} + G_{e_1,c}{}^{e_2} \hat{\overline{G}}^{e_1,}{}_{e_2}{}^c  - {1 \over 4} G_{e_1,e_2}{}^{e_2}  \hat{\overline{G}}^{e_1,c}{}_c  - 4 c_2 (\hat{G}^{e,c}{}_c \overline{G}_{e,d}{}^d + G_{e,}{}^{c}{}_c \hat{\overline{G}}^{e,}{}_{d}{}^d) ] \eqno(A.7) $$ 


\medskip
{\bf {References }}
\medskip
\item{[1]} P. West, {\it $E_{11}$ and M Theory}, Class. Quant. 
Grav.  {\bf 18}, (2001) 4443, hep-th/ 0104081.
\item{[2]} P. West, {\it $E_{11}$, SL(32) and Central Charges},
\item {[3]} A. Tumanov and P. West, {\it E11 must be a symmetry of strings and branes},  arXiv:1512.01644.
\item{[4]} A. Tumanov and P. West, {\it E11 in 11D}, Phys.Lett. B758 (2016) 278, arXiv:1601.03974.
\item{[5]} T. Curtright, {\it Generalised Gauge fields}, Phys. Lett. {\bf 165B} (1985) 304.
\item{[6]} C. Hull, {\it   Strongly Coupled Gravity and Duality}, Nucl.Phys. {\bf B583} (2000) 237, hep-th/0004195.
\item{[7]}  A. Tumanov and and P. West, {\it $E_{11}$,  Romans theory and higher level duality relations}, IJMPA, {\bf Vol 32}, No 26 (2017) 1750023,  arXiv:1611.03369
\item{[8]} A. Tumanov and and P. West, {\it  E11 and the non-linear dual graviton}, Phys.Lett. B779 (2018) 479-484, arXiv:1710.11031.
\item{[9]}  X. Bekaert, N. Boulanger and M. Henneaux, {\it Consistent deformations of dual formulations of linearized gravity: A no-go result } 
Phys.Rev. D67 (2003) 044010,  arXiv:hep-th/0210278. 
X.  Bekaert, N.  Boulanger and  S.  Cnockaert, {\it No Self-Interaction for Two-Column Massless Fields}, J.Math.Phys. 46 (2005) 012303, arXiv:hep-th/0407102. 
\item{[10]} P. West, {\it Introduction to Strings and Branes}, Cambridge University Press, 2012.
\item{[11]} P. West,{\it  A brief review of E theory}, Proceedings of Abdus Salam's 90th  Birthday meeting, 25-28 January 2016, NTU, Singapore, Editors L. Brink, M. Duff and K. Phua, World Scientific Publishing and IJMPA, {\bf Vol 31}, No 26 (2016) 1630043, arXiv:1609.06863. 
\item{[12]} P. West,  {\it On the different formulations of the E11 equations of motion}, Mod.Phys.Lett. A32 (2017) no.18, 1750096, arXiv:1704.00580.
\item{[13]} M. Gaberdiel,  D. Olive and P. West, {\it A class of Lorentzian Kac-Moody algebras} , Nucl.Phys. B645 (2002) 403-437,
hep-th/0205068.
\item{[14]} N. Lambert and P. West, {\it Coset Symmetries in Dimensionally Reduced Bosonic String Theory},  Nucl.Phys. B615 (2001) 117-132, hep-th/0107209.
\item{[15]} Very Extended $E_8$ and $A_8$ at low levels, Gravity and
Supergravity, Class.Quant.Grav. 20 (2003) 2393-2406, hep-th/0212291.
\item{[16]} A. Tumanov and P. West, {\it  Generalised vielbeins and non-linear realisations}, JHEP 1410 (2014) 009,  arXiv:1405.7894.
\item{[17]} M. Pettit and P. West,  {\it An E11 invariant gauge fixing}, Int.J.Mod.Phys. A33 (2018) no.01, 1850009, Int.J.Mod.Phys. A33 (2018) no.01, 1850009.
\item{[18]} A. Borisov and V. Ogievetsky, {\it Theory of
dynamical affine and conformal symmetries as gravity theory  of
 the gravitational field},  Theor.\ Math.\ Phys.\  {\bf 21} (1975)
1179; 
\item{[19]} V. Ogievetsky,  {\it ``Infinite-dimensional algebra ofgeneral  covariance group as the closure of the finite dimensional
algebras  of conformal and linear groups"},
Nuovo. Cimento, {\bf 8} (1973) 988.
\item{[20]} P. West, {\it Dual gravity and E11},  arXiv:1411.0920.
\item{[21]} P. West,  {\it   Generalised Space-time and Gauge Transformations }, arXiv:1403.6395.
\item{[22]}  M. Pettit and P. West,{\it E theory in seven dimensions},  Int.J.Mod.Phys. A34 (2019) no.25, 1950135,  arXiv:1905.07330.
\item{[23]}  F. Riccioni and P. West, {\it Dual fields and $E_{11}$},   Phys.Lett.
B645 (2007) 286-292,  hep-th/0612001
\item{[24]}  F. Riccioni,  D. Steele and P.West, {\it Duality Symmetries and $G^{+++}$ Theories},  Class.Quant.Grav.25:045012,2008,  arXiv0706.3659. 
\item{[25]}  G.  Barnich, P.  Mao and Romain Ruzziconi, {\it BMS current algebra in the context of the Newman-Penrose formalism}, arXiv:1910.14588
\item{[26]}  H.  Godazgar, M. Godazgar and C.N. Pope. {\it New dual gravitational charges},  arXiv:1812.01641; {it Dual gravitational charges and soft theorems} arXiv:1908.01164.
\item{[27]}  M. Pettit and P. West, {\it An E11 invariant gauge fixing},  Int.J.Mod.Phys. A33 (2018) no.01, 1850009, Int.J.Mod.Phys. A33 (2018) no.01, 1850009. 
\item{[28]} P. West, {Generalised BPS conditions}, Mod.Phys.Lett. A27 (2012) 1250202, arXiv:1208.3397. 
\item{[29]}  R,  Kallosh, D. Kastor, T.  Ortin and T.  Torma, {\it Supersymmetry and stationary solutions in dilaton-axion gravity},     Phys.Rev.D 50 (1994) 6374-6384, hep-th 9406059. 
\item{[30]} R. Argurio , F.  Dehouck and L. Houart,     Phys.Rev.D 79 (2009) 125001, hep-th 0810.4999.  
\item{[31]} P. West, {\it    Irreducible representations of E theory},  Int.J.Mod.Phys. A34 (2019) no.24, 1950133,  arXiv:1905.07324. 


\end